\documentclass[a4paper,fleqn]{cas-sc}
\usepackage{setspace}
\usepackage{longtable, booktabs}
\usepackage[numbers]{natbib}
\usepackage{xcolor}
\usepackage{xpatch}
\usepackage{siunitx}
\usepackage{caption}
\usepackage{etoolbox}
\usepackage{hyperref}

\hypersetup{
    colorlinks=true, 
    linkcolor=black, 
    filecolor=black, 
    urlcolor=black,  
    citecolor=black, 
}

\def\tsc#1{\csdef{#1}{\textsc{\lowercase{#1}}\xspace}}
\tsc{WGM}
\tsc{QE}
\tsc{EP}
\tsc{PMS}
\tsc{BEC}
\tsc{DE}

\begin{document}
\doublespacing

\sloppy

\hypersetup{citecolor=black}
\let\WriteBookmarks\relax
\def\floatpagepagefraction{1}
\def\textpagefraction{.001}
\tnotemark[1,2]

\author[1]{Cai}[prefix=Guohui]
\author[1]{Cai\texorpdfstring{\textsuperscript{}}{}}[prefix=Ying]\cormark[1]
\author[2]{Zhang}[prefix=Zeyu]\cormark[2]
\author[3]{Cao}[prefix=Yuanzhouhan]
\author[4]{Wu}[prefix=Lin]
\author[1]{Ergu}[prefix=Daji]
\author[5]{Liao}[prefix=Zhibin]
\author[6]{Zhao}[prefix=Yang]

\address[1]{College of Computer Science and Artificial Intelligence, Southwest Minzu University, Chengdu, 610225, China}
\address[2]{The Australian National University, Canberra ACT 2601, Australia}
\address[3]{School of Computer Science and Technology, Beijing Jiaotong University, Beijing, China}
\address[4]{Hunan university of technology, Zhuzhou, 410072, China}
\address[5]{University of Adelaide, Adelaide, South Australia 5005, Australia}
\address[6]{La Trobe University, Melbourne, Victoria 3086, Australia}
\cortext[1]{Corresponding author: Ying Cai (E-mail: caiying34@yeah.net).}
\cortext[2]{Project lead.}

\title{\LARGE Medical Artificial Intelligence for Early Detection of Lung Cancer: A Survey}
\shorttitle{Medical Artificial Intelligence for Early Detection of Lung Cancer: A Survey}

\shortauthors{Cai et al.}

\begin{abstract} 
Lung cancer remains one of the leading causes of morbidity and mortality worldwide, making early diagnosis critical for improving therapeutic outcomes and patient prognosis. Computer-aided diagnosis systems, which analyze computed tomography images, have proven effective in detecting and classifying pulmonary nodules, significantly enhancing the detection rate of early-stage lung cancer. Although traditional machine learning algorithms have been valuable, they exhibit limitations in handling complex sample data. The recent emergence of deep learning has revolutionized medical image analysis, driving substantial advancements in this field. This review focuses on recent progress in deep learning for pulmonary nodule detection, segmentation, and classification. Traditional machine learning methods, such as support vector machines and k-nearest neighbors, have shown limitations, paving the way for advanced approaches like Convolutional Neural Networks, Recurrent Neural Networks, and Generative Adversarial Networks. The integration of ensemble models and novel techniques is also discussed, emphasizing the latest developments in lung cancer diagnosis. Deep learning algorithms, combined with various analytical techniques, have markedly improved the accuracy and efficiency of pulmonary nodule analysis, surpassing traditional methods, particularly in nodule classification. Although challenges remain, continuous technological advancements are expected to further strengthen the role of deep learning in medical diagnostics, especially for early lung cancer detection and diagnosis. A comprehensive list of lung cancer detection models reviewed in this work is available at \url{https://github.com/CaiGuoHui123/Awesome-Lung-Cancer-Detection}.
\end{abstract}


\maketitle
\section{Introduction}
\begin{figure*}
  \begin{minipage}[t]{\linewidth}
    \centering
    \includegraphics[width=1\textwidth]{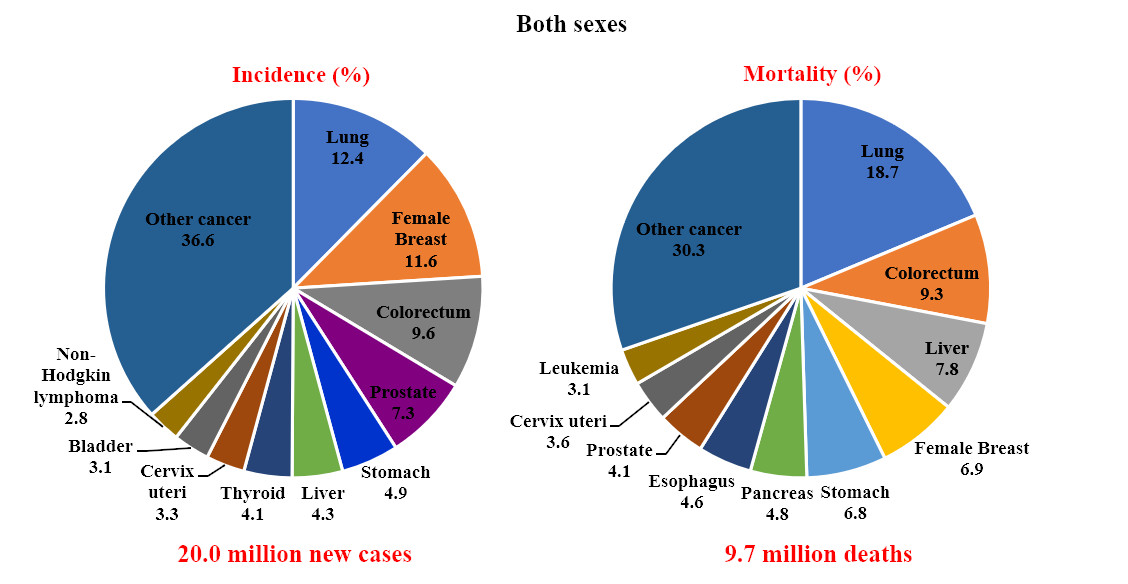}
    \caption{Pie chart showing the distribution of cases and deaths in 2022 for the top five cancers and an aggregated “Others” category (all remaining cancer types). The area of each slice reflects its proportion of the total number of cases or deaths \cite{bray2024global}.}
    \label{fig:1}
  \end{minipage}
\end{figure*}

\doublespacing
Lung cancer is one of the leading causes of both incidence and mortality worldwide. According to the latest assessment by the International Agency for Research on Cancer (IARC) of the World Health Organization (WHO), in 2022, there were approximately 19.96 million new cancer cases globally, with 2.48 million of them being lung cancer, accounting for 12.4\% of all cases, making it the most prevalent cancer worldwide {\hyperref[fig:1]{Fig. \ref*{fig:1}}} \cite{bray2024global}. Due to the subtlety of early-stage lung cancer symptoms, patients often miss the optimal treatment window. Early screening and computed tomography (CT) scans are essential for early diagnosis, as they allow for the detection of lesions before clinical symptoms manifest, significantly increasing the rate of early lung cancer detection. Computer-aided diagnosis (CAD) systems are designed to assist in medical image analysis by automating the detection of abnormalities, such as those in CT scans \cite{jin2023machine}. These systems improve diagnostic accuracy and reduce the workload of radiologists. In lung cancer detection, CAD systems are particularly effective in identifying and assessing pulmonary nodules. With advances in deep learning and artificial intelligence, these systems have become increasingly accurate, fast, and reliable, supporting early lung cancer diagnosis.

\begin{figure*}
  \begin{minipage}[t]{\linewidth}
    \centering
    \includegraphics[width=1\textwidth]{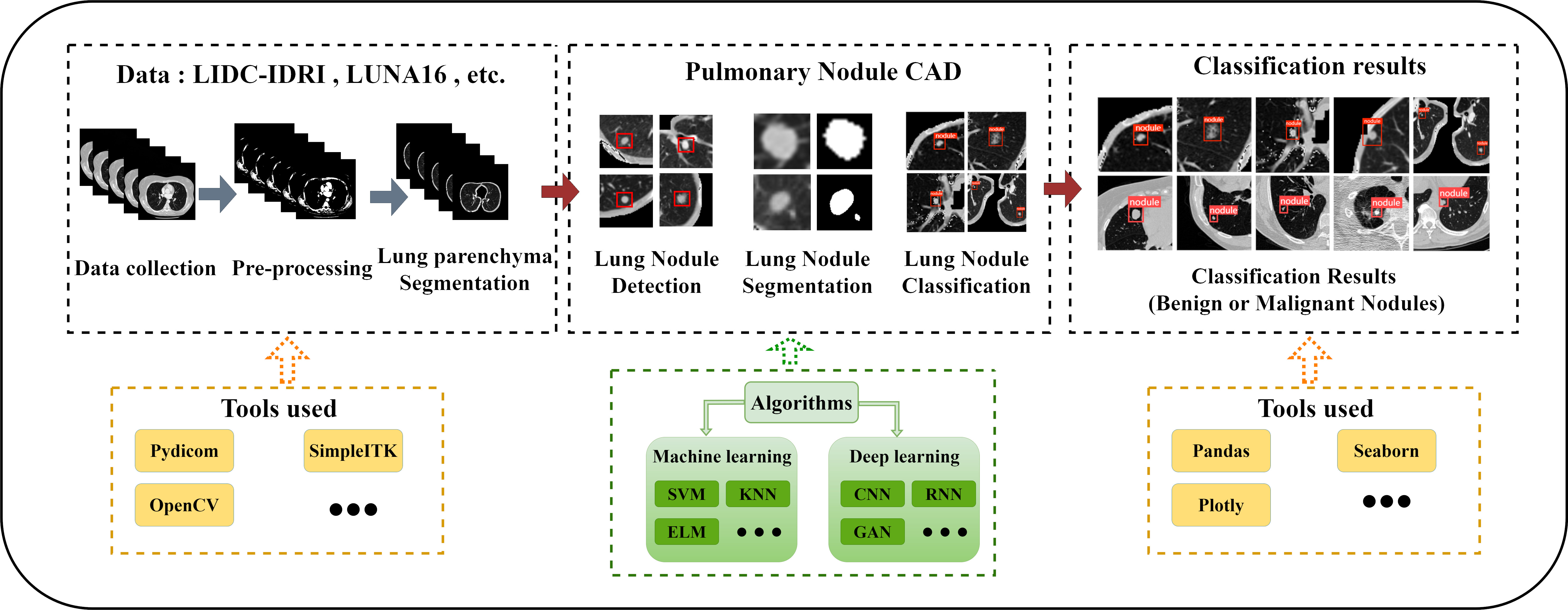}
    \caption{Framework of this study and the basic steps of the pulmonary nodule CAD system.}
    \label{fig:2}
  \end{minipage}
\end{figure*}

Pulmonary nodules are defined as round or irregular lesions with a diameter of less than 30 mm, typically appearing on CT images as high-density areas with either well-defined or blurred margins. Nodules can be categorized into solid, subsolid, and ground-glass types based on their density and morphology. Among these, solid nodules generally pose a lower risk of malignancy, whereas subsolid and ground-glass nodules, particularly those that are large and irregular, are associated with a higher risk. Consequently, accurate identification and classification are critical for clinical diagnosis \cite{zhang2024deep,hiwase2024can}. The application of computer-aided detection (CAD) technology in lung cancer diagnosis began in the 1980s, initially focusing on assisting in the detection of abnormalities on chest X-rays. It later expanded to high-resolution CT image analysis, evolving from simple image processing to sophisticated deep learning models that significantly enhance both sensitivity and specificity. Today, CAD systems can autonomously detect and evaluate pulmonary nodules, provide follow-up recommendations, and assist in decision-making, substantially improving the accuracy and efficiency of early lung cancer screening.

Compared to traditional machine learning (ML) methods, deep learning, particularly convolutional neural networks (CNNs), has demonstrated superior performance in pulmonary nodule CAD systems \cite{krizhevsky2012imagenet}. Traditional ML relies on manually extracted features, which are limited by the quality of the features, whereas deep learning can automatically learn complex image features, significantly improving detection accuracy and robustness. This advantage is especially apparent in handling high-dimensional and large-scale data, where deep learning excels at capturing subtle changes in pulmonary nodules, thereby enhancing diagnostic efficiency. The workflow of a pulmonary nodule CAD system includes CT image preprocessing, automatic segmentation of nodule regions, feature extraction, and classification analysis, ultimately generating diagnostic reports that assist clinicians in decision-making. The system aims to improve detection accuracy and efficiency, contributing to early lung cancer diagnosis.

\begin{figure*}
  \begin{minipage}[t]{\linewidth}
    \centering
    \includegraphics[width=1\textwidth]{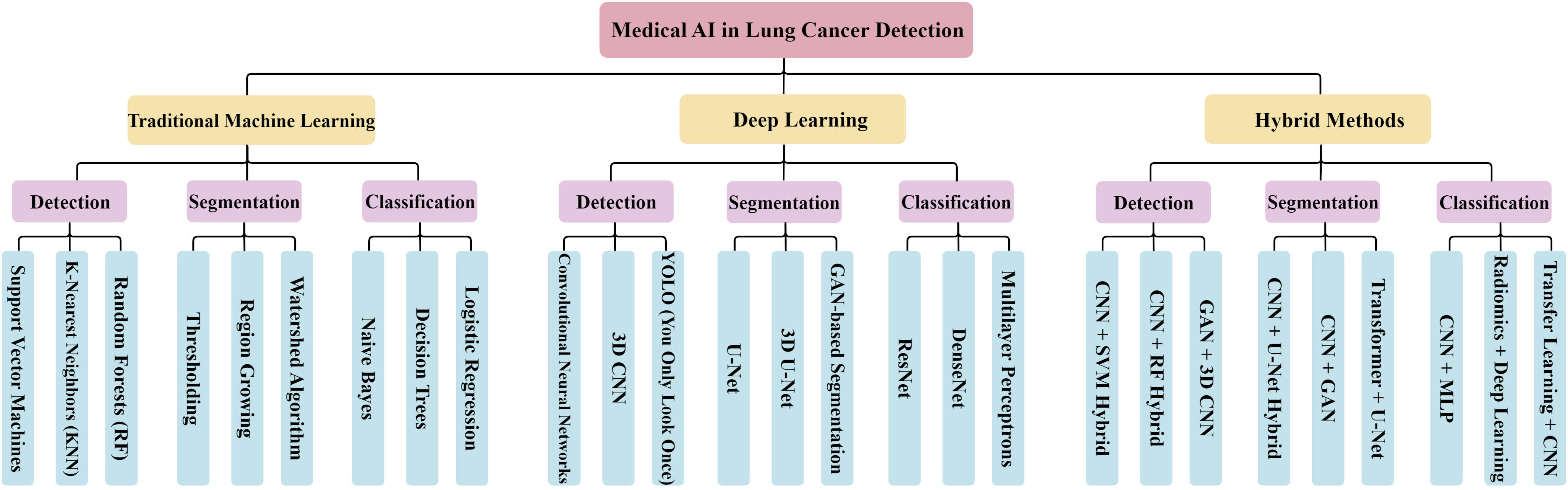}
    \caption{This tree diagram illustrates the overall development of medical AI technologies in the early detection of lung cancer. The diagram provides an overview of key advancements and techniques in the field, structured into three main sections: Traditional Machine Learning, Deep Learning, and Hybrid Methods. Each section is further divided into three core tasks—Detection, Segmentation, and Classification—highlighting key techniques used in each. Traditional methods like Support Vector Machines (SVM), Thresholding, and Decision Trees have been foundational, while deep learning methods such as CNN, U-Net, and ResNet have significantly advanced the field, especially after the rise of deep learning in 2012. Hybrid methods, combining models from both machine learning and deep learning, have become prominent since 2018, offering enhanced accuracy and robustness.}
    \label{fig:3}
  \end{minipage}
\end{figure*}

Recent reviews have summarized the application of AI in early lung cancer detection. The study by Kaulgud et al. \cite{kaulgud2023analysis} primarily focuses on CT image detection methods and compares the performance of various techniques in pulmonary nodule detection. Similarly, Kalkeseetharaman et al. \cite{kalkeseetharaman2024bird} review studies based on X-ray and CT images, highlighting the advantages of deep learning in CT image-based pulmonary nodule detection. However, these reviews lack an in-depth exploration of the differences between traditional machine learning and deep learning algorithms, particularly in terms of data sources and nodule segmentation methods, which is a significant oversight \cite{kaulgud2023analysis}. They also provide insufficient descriptions of specific algorithms and model characteristics, lacking a unified review framework \cite{kalkeseetharaman2024bird}.

To address these gaps, this review discusses the application of AI technologies in early lung cancer detection and diagnosis, aiming to leverage AI for the rapid detection \cite{zhao2024landmark}, segmentation \cite{ge2024esa}, and classification of pulmonary nodules. The research framework is illustrated in {\hyperref[fig:2]{Fig. \ref*{fig:2}}}, and the overall development of medical AI technologies in lung cancer detection is shown in {\hyperref[fig:3]{Fig. \ref*{fig:3}}}.

The main contributions of this review are as follows:
\begin{itemize} 
\item A comprehensive analysis of diverse data sources and a systematic literature retrieval and screening process, identifying key studies and trends in lung cancer detection. 
\item A discussion of the structure, performance, and applicable scenarios of traditional machine learning and deep learning algorithms for lung cancer analysis. 
\item An in-depth comparison of lung cancer detection, segmentation, and classification tasks across 131 reviewed studies, highlighting their respective strengths, limitations, and future research directions. 
\item Practical insights into designing high-performance lung cancer analysis workflows, optimizing computational complexity, and enhancing model interpretability, providing valuable guidance for researchers and radiologists. 
\end{itemize}

\section{Medical AI for Lung Cancer Detection}
In recent years, the rapid advancement of artificial intelligence has brought unprecedented transformations across various industries, particularly in the field of lung cancer detection and diagnosis. Traditional methods for detecting pulmonary nodules often rely on the expertise of radiologists, are time-consuming, and are influenced by subjective factors, which can compromise the accuracy of diagnostic results. However, with the progress of AI technology, the healthcare sector has embraced new possibilities. AI enables machines and computer systems to emulate human intelligence, encompassing capabilities such as language comprehension, natural language processing, decision-making, and visual perception, as discussed by Messeri et al. \cite{bawack2021framework}. The robust learning capabilities and high precision of AI have led to its increasing application in lung cancer detection, particularly in the analysis of medical images. Through deep learning algorithms, AI can automatically identify and analyze pulmonary nodules, significantly enhancing the accuracy of early diagnosis and assisting physicians in making timely and effective decisions, ultimately improving patient survival rates.

\subsection{Statistical Learning Based}
\begin{figure*}
  \begin{minipage}[t]{\linewidth}
    \centering
    \includegraphics[width=1\textwidth]{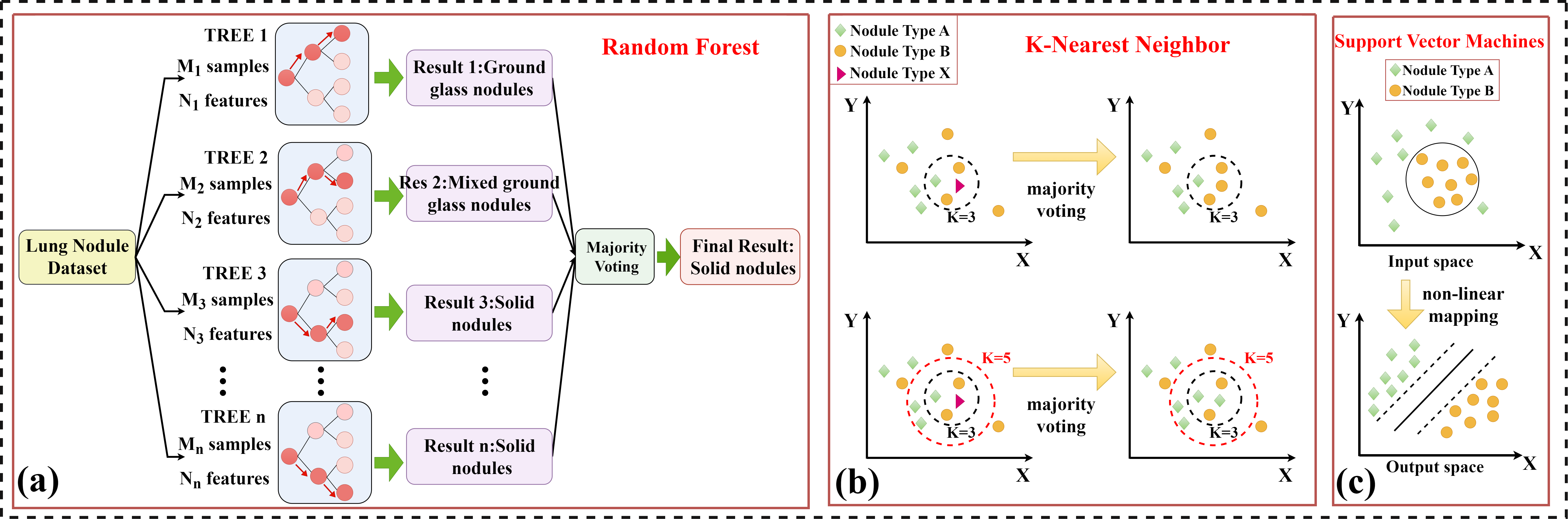}
    \caption{Algorithm structure schematic. (a) Random Forest (RF) incorporates multiple decision trees to evaluate different aspects of the pulmonary nodule dataset, with a majority vote determining the final category of each nodule. (b) K-Nearest Neighbor (KNN) classifies nodules by analyzing the closest training examples in the feature space, employing a specific number of neighbors (e.g., K=3) for voting. (c) Support Vector Machine (SVM) constructs a hyperplane to distinctly separate different types of nodules, optimizing classification based on their features. Each method effectively leverages feature information along the X and Y axes to distinguish between nodule types.}
    \label{fig:4}
  \end{minipage}
\end{figure*}

Machine learning is a methodology for implementing artificial intelligence that utilizes statistical learning algorithms to enable computer systems to learn from experience and improve performance. In this process, manually designed feature engineering is crucial for transforming data into a format suitable for shallow models. As illustrated in {\hyperref[fig:4]{Fig. \ref*{fig:4}}}, statistical learning-based algorithms include Support Vector Machines (SVM) \cite{cortes1995support}, K-Nearest Neighbors (KNN) \cite{cover1967nearest}, Extreme Learning Machines (ELM) \cite{huang2006extreme}, Decision Trees (DT), Random Forests (RF), and Naive Bayes (NB). These methods have been widely applied in medical imaging analysis, particularly in the detection of pulmonary nodules, significantly improving diagnostic accuracy and assisting physicians in identifying potential malignant lesions in a timely manner.

\begin{figure*}
  \begin{minipage}[t]{\linewidth}
    \centering
    \includegraphics[width=1\textwidth]{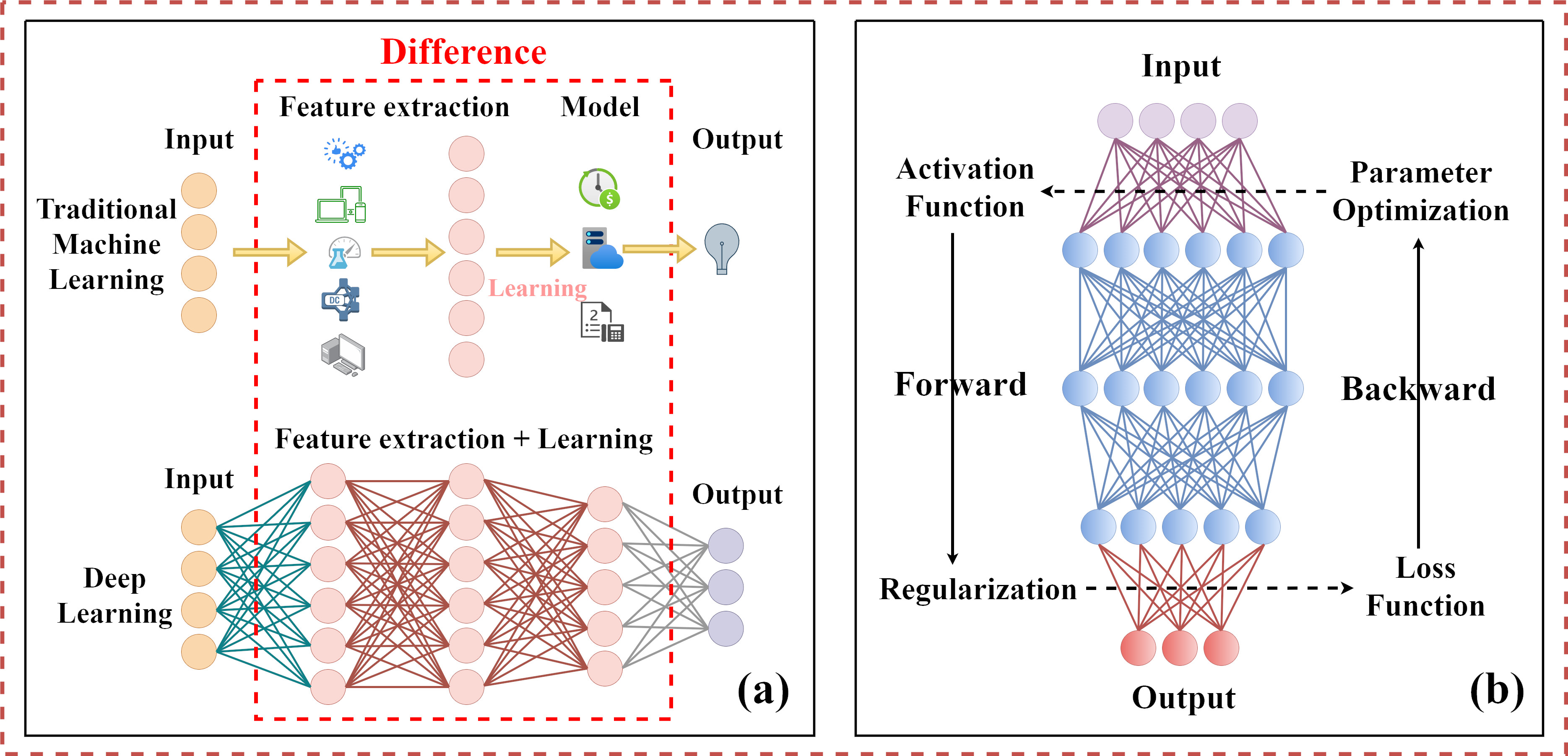}
    \caption{(a) Main difference between statistical learning-based and deep learning. Statistical learning-based methods rely on manually extracted features and employ simpler algorithms that often require human intervention for optimal performance. In contrast, deep learning automates the feature extraction process, utilizing complex neural network architectures that learn directly from raw data. This capability enables deep learning to handle more complex data structures and patterns, improving accuracy and efficiency in tasks such as medical image analysis. (b) The deep neural network operation process.}
    \label{fig:5}
  \end{minipage}
\end{figure*}

Despite their commendable performance in this field, statistical learning-based methods have certain limitations. They often rely on manual feature design and selection, which can be time-consuming and heavily influenced by domain knowledge; improper feature selection may adversely affect model performance. Additionally, these algorithms face challenges in effectively handling nonlinearity, high-dimensional data, and complex data relationships, particularly when dealing with unstructured data such as lung imaging, where capturing latent patterns and structures proves difficult. However, the advent of deep learning methods offers effective solutions to these challenges. Deep learning has significant advantages over statistical learning-based methods, with the main differences shown in {\hyperref[fig:5]{Fig. \ref*{fig:5}}}a. Deep learning models can autonomously learn feature representations from raw data, eliminating the need for tedious manual feature engineering. The operation process of a deep neural network is illustrated in {\hyperref[fig:5]{Fig. \ref*{fig:5}}}b.

\subsection{Convolutional Neural Network (CNN)}
\begin{figure*}
  \begin{minipage}[t]{\linewidth}
    \centering
    \includegraphics[width=1\textwidth]{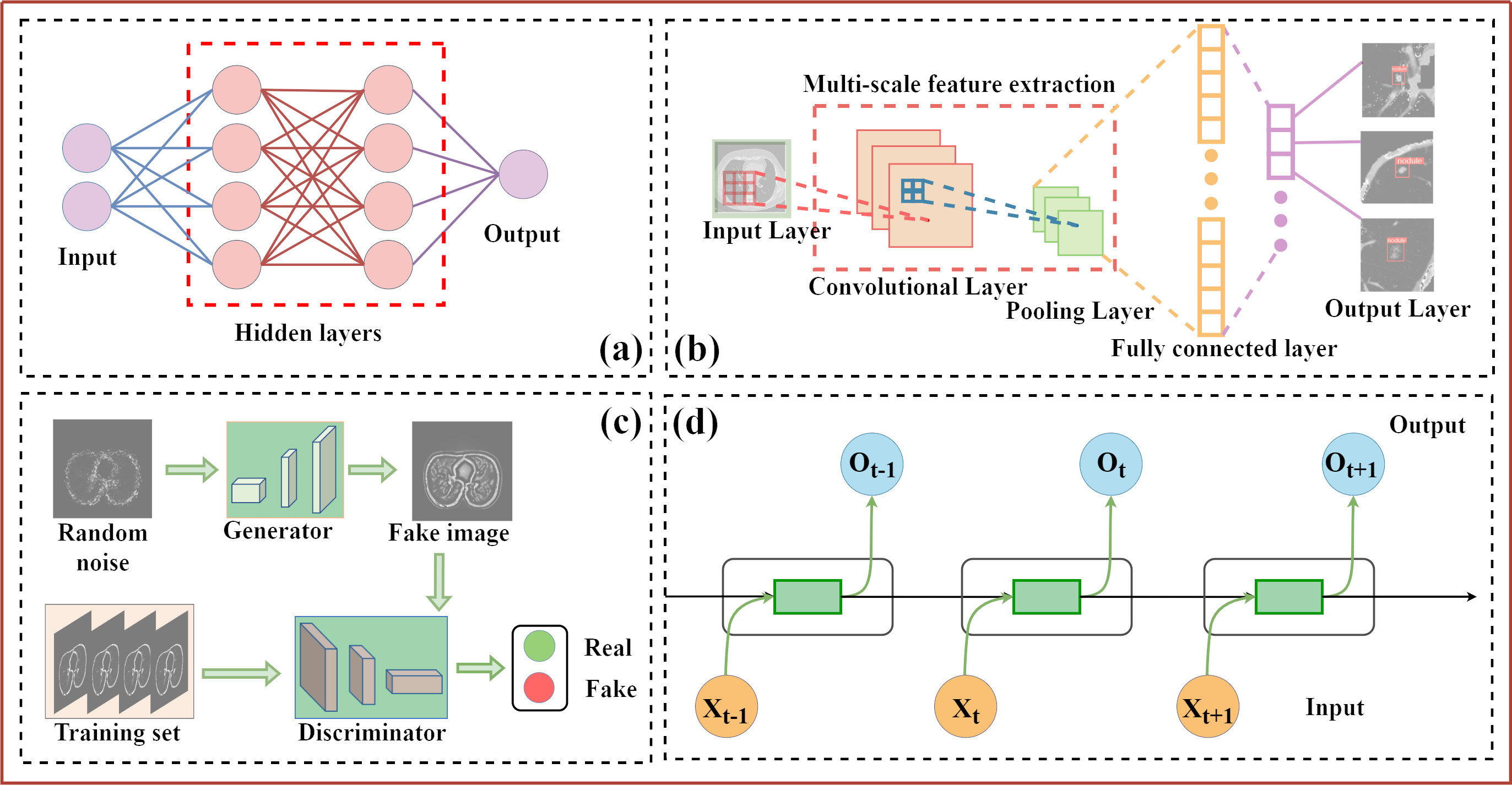}
    \caption{Various deep learning models and structures: (a) Multilayer Perceptron (MLP) performs complex function approximation, vital for pattern recognition in data structures. (b) Convolutional Neural Network (CNN) excels in automatic feature extraction and analyzing structural details in lung CT scans. (c) Generative Adversarial Network (GAN) synthesizes accurate medical images, mitigating the lack of annotated samples and improving model robustness. (d) Recurrent Neural Network (RNN) analyzes temporal patterns in sequential lung scans, crucial for accurate pulmonary nodule detection.}
    \label{fig:6} 
  \end{minipage}
\end{figure*}

Convolutional Neural Networks (CNNs) originated in the 1980s with the development of LeNet by Yann LeCun's team for handwritten digit recognition \cite{lecun1998gradient}. The rapid advancements in computational power and large-scale datasets facilitated the widespread adoption of CNNs in the early 21st century, culminating in the success of AlexNet in the 2012 ImageNet competition, which marked a significant milestone \cite{krizhevsky2012imagenet}. Subsequent architectures like DenseNet and ResNet further improved classification accuracy \cite{he2016deep}, significantly advancing both computer vision and medical image analysis. Today, CNNs have become the cornerstone of modern medical image analysis, particularly in the detection \cite{zhang2024meddet}, segmentation \cite{wu2023bhsd,zhang2024segreg,zhang2023thinthick}, and classification of pulmonary nodules in CT scans. Designed to automatically and adaptively learn spatial hierarchies of features from input images through multiple layers of convolutional filters, CNNs capture intricate patterns and textures inherent in medical imaging data, making them exceptionally well-suited for identifying subtle abnormalities such as pulmonary nodules. Fundamental architectures such as LeNet, AlexNet, VGG, ResNet, and, more recently, specialized models like U-Net and DenseNet, have been extensively employed and further refined to enhance the precision and efficiency of lung cancer detection systems. The basic structure of a CNN is shown in {\hyperref[fig:6]{Fig. \ref*{fig:6}}}b.

In addition to CNNs, Multilayer Perceptrons (MLPs) have historically served as foundational models in deep learning, consisting of an input layer, multiple hidden layers, and an output layer, as shown in {\hyperref[fig:6]{Fig. \ref*{fig:6}}}a. While MLPs excel at capturing higher-order abstract features within data, their application to complex, high-dimensional tasks like pulmonary nodule classification often leads to increased computational costs and challenges such as overfitting. To overcome these limitations, researchers have integrated MLPs with CNN architectures, leveraging the strengths of both models. For instance, Zhu et al. \cite{zhu2018deeplung} enhanced nodule detection by combining convolutional layers with MLP components in CM-Net, effectively capturing local dependencies in CT images while maintaining computational efficiency.

\begin{table*}
\scriptsize %
\setlength{\tabcolsep}{0.205cm} %
\renewcommand{\arraystretch}{1} %
\setlength{\heavyrulewidth}{1pt} %
    \centering 
   \caption{Typical applications of CNN models in pulmonary nodule detection, segmentation, and classification on CT images.} 
\label{table:1}
\begin{tabular}{@{\extracolsep{\fill}} p{0.6cm} | p{1.5cm} p{0.5cm} p{0.7cm} p{1.6cm} p{1.6cm} p{1.8cm} p{1.9cm} p{2.9cm} @{}}
\toprule
  \makecell[l]{Task\\Type} & \makecell[l]{Authors} & \makecell[l]{Year} & \makecell[l]{2D/3D} & \makecell[l]{Technique/\\Network Type} & \makecell[l]{Validation/\\Test Method} & \makecell[l]{Nodules Types} & \makecell[l]{Main Method} & \makecell[l]{Model Performance} \\
  \midrule
\specialrule{0.1pt}{0pt}{0pt} 
& Kadhim et al. {\cite{kadhim2022computer}} & 2022 & 2D and 3D & CNN ; SAE & Accuracy, Sensitivity & Multiple appearances and shapes (size <10mm) & Pulmonary nodule detection and diagnosis with CAD & Performance varies based on dataset used, comparison of CAD methods \\
& Bhatt et al. {\cite{bhatt2022diagnosis}} & 2022 & 2D & YOLOv4 & Sensitivity & Various nodule types & YOLOv4 model & Sensitivity with FP/scan: 81\% \\
& Mesquita et al. {\cite{de2022novel}} & 2022 & 3D & CNN & 10-fold cross-validation & Nodules larger than 3 mm & Nodule detection using 16 filters and Boolean logic & Sensitivity: 92.75\%, 8 false positives per exam with majority gold standard \\
& Suzuki et al. {\cite{suzuki2022development}} & 2022 & 3D & CNN;3D U-Net & Sensitivity & Nodules > 5 mm & Automated nodule detection with 3D U-Net & Internal CPM: 94.7\%, External CPM: 83.3\% \\
Det-ection & Karrar et al. {\cite{karrar2023auto}} & 2022 & 2D and 3D & SVM, DCNN & 10-fold cross-validation & Solitary, juxtapleural & CADx: segmentation, SVM, DCNN & SVM: 91.4\%, DCNN: 95\% \\
& Bhaskar et al. {\cite{bhaskar2023pulmonary}} & 2023 & 3D & U-net, CNN & 80:20 train-test split & Cancerous or non-cancerous & U-net for segmentation, CNN for classification & Sensitivity: 0.75 (before), 0.65 (after classification) \\
& Jian et al. {\cite{jian2024dbpndnet}} & 2024 & 3D & Dual-branch, 3D CNN & 5-fold cross-validation & Lung parenchyma and chest wall nodules & Multi-task dual-branch 3D CNN, attention fusion & Sensitivity: 91.33\%, 0.125 to 8 FPs/scan \\
& Zhao et al. {\cite{zhao2023attentive}} & 2023 & 3D & 3DCNN & 10-fold cross-validation & Types, shapes, sizes: 3-30 mm & Adaptive 3D CNN & Sensitivity: 0.947, FP/s: 0.14 \\
& Ahmadyar et al. {\cite{ahmadyar2024hierarchical}} & 2024 & 2D and 3D & YOLOv5s, 3DCNN & Evaluated on LUNA16 & Pulmonary nodules & Hierarchical YOLOv5s, 3DCNN & Sensitivity: 97.8\%, confidence: 0.3 \\
& Ma et al. {\cite{ma2024ticnet}} & 2024 & 3D & Transformer, CNN & 10-fold cross-validation & Benign and malignant & TiCNet, Transformer, attention, multi-scale fusion & Sensitivity: 93\%, <11 FP/scan, CPM: 90.73\% \\
\midrule
& Sweetline et al. {\cite{sweetline2024overcoming}} & 2024 & 2D & Multi-crop CNN & Sensitivity & Various nodule types & Multi-crop CNN, boundary refinement, seg & DSC:LUNA16:0.978, LIDC:0.982, Sensitivity: LUNA16:97.6\%, LIDC:98\% \\
& Zhang et al. {\cite{zhang2024ds}} & 2024 & 2D & DS-MSFF-Net & Sensitivity & Benign and malignant & Dual-path: semantic feature extraction & LIDC-IDRI: 85.39\%, LiTS2017 liver: 95.79\%, LiTS2017 tumor: 91.75\% \\
Segme-ntation & Suji et al. {\cite{suji2024exploring}} & 2024 & 2D & UNet, FPN, PSPNet & Sensitivity & Various nodule types & Pretrained encoders: ResNet & DSC: IoU (UNet-efficientnet-b3): 0.5922 \\ 
& Asiya et al. {\cite{asiya2024automatically}} & 2024 & 2D & Custom-VGG16 & Sensitivity & Benign and malignant & Custom-VGG16: preprocessing & Precision: 90.87\% \\
& Tang et al. {\cite{tang2023sm}} & 2023 & 2D & SM-RNet & Sensitivity & Benign and malignant & SM-RNet: Weighted-fusion & Precision: FUSCC: 89.774\%, LUNA16: 85.047\% \\
& Cai et al. {\cite{cai2024mdfn}} & 2024 & 2D & MDFN & 5-fold cross-validation & Various nodule types & MDFN: Self-calibrated edge enhancement & DSC: 89.19\%; Sensitivity: 88.89\% \\
\midrule
& Fu et al. {\cite{fu2021fusion}} & 2021 & 3D & MLP ; Mr-Mc & 5-fold cross-validation & Inflammation, squamous cell carcinoma & Double modal fusion of CT images and LTBs & Mr-Mc Acc: 0.810, MLP Acc: 0.887, Fusion model Acc: 0.906 \\
& Zhan et al. {\cite{zhan2023uncertainty}} & 2023 & 3D & 3D CNN: SSL & 5-fold cross-validation & Solitary Pulmonary Nodules (SPNs) & SSL with USC for supervised training & Supervised: Acc: 0.662, USC (confidence): Acc: 0.702, Sensitivity: 0.707 \\
Class-ification & Rahouma et al. {\cite{rahouma2024automated}} & 2024 & 3D & 3D CNN + GA & 4-fold cross-validation & Benign or Malignant & 3D CNN with GA for design & Acc: 95.98\%, Sensitivity: 98.80\%, AUC: 0.985, Specificity: 93.40\% \\
& Lin et al. {\cite{lin2024combined}} & 2024 & 3D & 3D CNN & 10-fold cross-validation & IA, MIA, AIS, and AAH & AutoGluon-Tabular for classification & Task 1: Acc: 92.8\%, Sensitivity: 89.43\%, AUC: 96.17\%, Task 2.1: Acc: 74.76\% \\
& Drishti et al. {\cite{drishti2024novel}} & 2024 & 2D & CNN - ConvNet & 5-fold cross-validation & benign or Malignant & Multi-scale, multi-path for feature extraction & Acc: 90.38\%, Sensitivity: 88.70\%, AUC: 0.948, Specificity: 92.40\% \\
  \bottomrule
\end{tabular} 
\end{table*}

When comparing various CNN-based approaches for pulmonary nodule analysis, several studies demonstrate significant advancements in both accuracy and computational efficiency. For instance, Kadhim et al. \cite{kadhim2022computer} developed a high-efficiency CAD system using CNNs and stacked autoencoders, achieving an accuracy of 99.0\% on the LIDC-IDRI dataset. Similarly, Bhatt et al. \cite{bhatt2022diagnosis} employed the YOLOv4 model, which attained 95\% precision and 81\% sensitivity, showcasing real-time detection capabilities. Mesquita et al. \cite{de2022novel} introduced a Boolean equation-based method integrated with CNNs, achieving 92.75\% sensitivity with 8 false positives per scan. Additionally, Suzuki et al. \cite{suzuki2022development} enhanced a 3D U-Net model, reaching a CPM of 94.7\% on the LIDC-IDRI dataset and 83.3\% on a Japanese dataset, demonstrating robust cross-dataset performance. These studies highlight the versatility and robustness of CNN architectures in diverse clinical settings, effectively handling different nodule types and minimizing false positives.

Furthermore, the integration of MLPs within CNN frameworks has led to the development of more sophisticated models that combine feature extraction with complex classification tasks. For example, Fu et al. \cite{fu2021fusion} introduced a multimodal diagnostic model that integrates 3D lung CT images with serum biomarkers, achieving an average accuracy of 90.6\% on the LIDC-CISB dataset by fusing a multi-resolution 3D multi-class deep learning model (Mr-Mc) with an MLP model based on lung tumor biomarkers (LTBs). This integration underscores the potential of hybrid models in enhancing the accuracy and reliability of pulmonary nodule detection and classification.

In summary, CNNs are pivotal to modern computer-aided diagnosis (CAD) systems for lung cancer, offering unmatched capabilities in feature extraction and pattern recognition. Models such as Kadhim et al.'s \cite{kadhim2022computer} CNN with stacked autoencoders, Bhatt et al.'s \cite{bhatt2022diagnosis} YOLOv4, de Mesquita et al.'s \cite{de2022novel} Boolean equation-based CNN, Suzuki et al.'s \cite{suzuki2022development} enhanced 3D U-Net, and the MLP-integrated approaches by Zhu et al. \cite{zhu2018deeplung} and Fu et al. \cite{fu2021fusion} demonstrate significant strides in improving the accuracy, sensitivity, and specificity of pulmonary nodule detection and classification. These advancements not only facilitate early diagnosis and optimize treatment planning but also support the seamless integration of CNN-based systems into routine clinical workflows, thereby elevating the standards of lung cancer care. The typical CNN applications in this field in recent years are listed in Table ~\ref{table:1}.

\subsection{Recurrent Neural Network (RNN) }
\begin{figure*}
  \begin{minipage}[t]{\linewidth}
    \centering
    \includegraphics[width=1\textwidth]{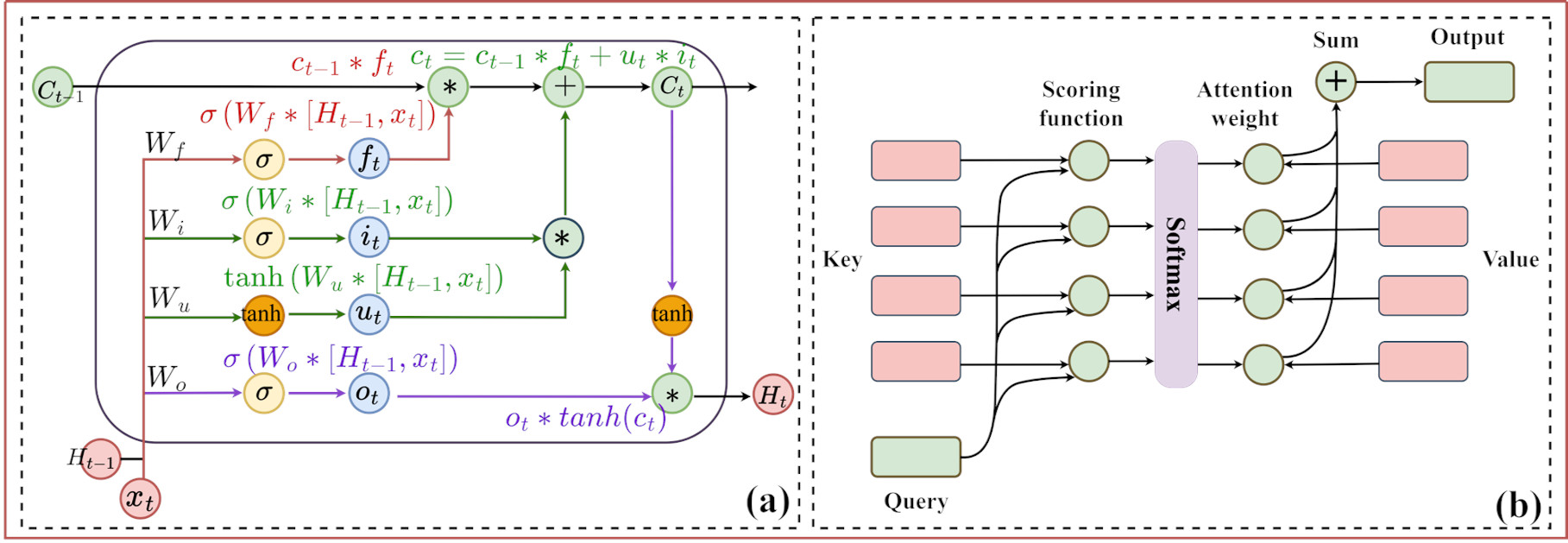}
    \caption{Various deep learning models and structures: (a) Long Short-Term Memory (LSTM) structure is vital for sequence data analysis in medical applications, utilizing forget, input, and output gates to manage information flow, thereby maintaining essential temporal features critical for accurate diagnostic predictions. This mechanism proves especially effective in analyzing diagnostic time-series data, capturing crucial changes over time. (b) The Attention Mechanism enhances neural network focus on significant features within diagnostic images, facilitating precise detection and categorization of anomalies such as tumors and lesions by adjusting the model's focus to the most informative regions of the image data.}
    \label{fig:7}
  \end{minipage}
\end{figure*}

Recurrent Neural Networks (RNNs), introduced by Rumelhart et al. in 1986 \cite{rumelhart1986learning}, are designed to handle sequential data by maintaining an internal state that captures information from previous inputs. This makes RNNs particularly effective for tasks involving temporal dependencies, such as language processing and the detection of pulmonary nodules from continuous imaging datasets. Traditional RNNs, however, face challenges like vanishing and exploding gradients, limiting their ability to learn long-term dependencies. To address these issues, architectures such as Long Short-Term Memory (LSTM) networks and Gated Recurrent Units (GRUs) were developed \cite{hochreiter1997long}, employing gating mechanisms to control the flow of information and retain context over longer sequences. In pulmonary nodule detection, these models improve accuracy by learning inter-slice dependencies in CT images. Furthermore, Bidirectional RNNs and attention mechanisms enhance the network's ability to focus on important parts of the sequence, increasing detection precision. With these advancements, RNNs and their variants have become essential tools in medical image analysis, particularly in the early detection of lung cancer, as illustrated in {\hyperref[fig:6]{Fig. \ref*{fig:6}}}d and {\hyperref[fig:7]{Fig. \ref*{fig:7}}}a.

In the analysis of pulmonary nodules, RNNs, with their unique ability to process temporal sequences, are capable of capturing sequential patterns in CT scan data, significantly improving classification tasks. For instance, Balannolla et al. \cite{balannolla2022detection} combined RNNs with convolutional layers for pulmonary nodule classification, achieving a sensitivity of 96.4\% using KNN classifiers. Similarly, Vijay et al. \cite{gugulothu2024early} employed an RNN model integrated with feature selection algorithms for nodule classification, reaching a classification accuracy of 93.6\% and a sensitivity of 96.39\%. The advantage of RNNs in these medical imaging tasks lies in their ability to handle temporal dependencies within sequential data, thereby effectively improving the robustness and accuracy of classification. However, the complexity and high computational demands of RNNs remain challenges that need optimization, with researchers continuously exploring novel network architectures and optimization strategies to address these issues.

Although significant progress has been made in pulmonary nodule detection using convolutional neural networks (CNNs) in recent years, RNNs, as a deep learning method capable of processing temporal information, continue to show potential in enhancing performance when integrated with other network architectures. For example, some studies have combined RNNs with CNNs to improve nodule detection, offering the dual advantage of handling both spatial and temporal features. Additionally, methods such as the adaptive anchor box Faster R-CNN proposed by Nguyen et al. \cite{nguyen2023manet}, which integrates RNNs, and the dual-stage framework developed by Shimaa et al. \cite{el2020two}, where RNN modules enhance classification tasks, have made significant progress in improving detection sensitivity and reducing false positives. These studies illustrate that the application of RNNs in medical imaging extends beyond classification tasks, and when combined with other neural networks, can further enhance the recognition and handling of complex data patterns, thereby advancing the field of pulmonary nodule detection.

\subsection{Generative Adversarial Network (GAN)}
Generative Adversarial Networks (GANs), introduced by Ian Goodfellow and colleagues in 2014 \cite{goodfellow2014generative}, consist of two neural networks—the generator and the discriminator—that are trained simultaneously through an adversarial process. The basic structure is shown in {\hyperref[fig:6]{Fig. \ref*{fig:6}}}c. The generator aims to produce realistic data samples, while the discriminator attempts to distinguish between genuine data and those generated by the generator. This competitive dynamic drives the generator to produce increasingly realistic data, enhancing its capacity to model complex data distributions. In medical imaging, GANs have been extensively applied to tasks such as image synthesis, restoration, and super-resolution, significantly improving the accuracy and robustness of pulmonary nodule detection. Variants of GANs, such as Conditional GANs, Generative Adversarial Autoencoders, and CycleGANs, provide specialized solutions for specific challenges. For example, Conditional GANs generate images with specific pathological features, increasing data diversity, while Generative Adversarial Autoencoders produce high-resolution images that assist in the precise identification of nodules. CycleGANs, notably, can improve image quality without the need for paired datasets, which is particularly beneficial in medical scenarios with limited annotated data. By generating high-fidelity synthetic data for data augmentation, GANs help address issues of limited and imbalanced datasets, resulting in enhanced model performance and generalization. Overall, GANs and their variants play a crucial role in enriching diagnostic data and driving advancements in medical image analysis, especially for the early detection of pulmonary nodules.

In the realm of pulmonary nodule analysis, GANs have been employed to tackle challenges related to data scarcity and class imbalance. For instance, Sengodan et al. \cite{sengodan2023early} applied GANs to generate synthetic CT images of pulmonary nodules, augmenting the training dataset and improving the robustness of detection models. In another study, Ling et al. \cite{zhu2021hr} incorporated GAN-based data augmentation into a dual-branch CNN framework, leading to improved segmentation accuracy and reduced false positives. Furthermore, Gugulothu et al. \cite{gugulothu2023novel} introduced a deep learning method that integrates step deviation mean multi-level thresholding (SDMMT) and a novel CSDR-J-WHGAN classifier, achieving impressive classification results with 97.11\% detection accuracy on the LIDC-IDRI dataset. Zhu et al. \cite{zhu2021functional} further enhanced GAN-based approaches by developing the Functional Realistic GAN (FRGAN) architecture, which employs TreeGAN to generate high-resolution images with accurate pathological features, significantly improving detection sensitivity and accuracy, as demonstrated by its CPM score of 0.915 on the LUNA16 dataset.

Moreover, Tyagi et al. \cite{tyagi2022cse} proposed a 3D Conditional GAN (CSE-GAN) for pulmonary nodule segmentation, integrating a U-Net-based generator with parallel compression and excitation modules. This approach effectively reduced overfitting and improved segmentation performance, achieving a Dice coefficient of 80.74\% on the LUNA16 dataset. These advanced GAN-based methods not only enhance training datasets with high-quality synthetic images but also improve segmentation and classification accuracy, effectively addressing issues related to data scarcity and class imbalance. On the other hand, we have greatly enriched our discussion of Generative Adversarial Networks (GANs) by detailing not only their applications in data augmentation but also the quantitative metrics used to assess the quality and realism of synthetic medical images. Key evaluation measures now include the Fréchet Inception Distance (FID), which quantifies distributional similarity between generated and real images in the feature space of a pretrained network (lower is better); the Inception Score (IS), which captures both image quality and diversity via entropy-based measures (higher is better); the Structural Similarity Index Measure (SSIM), commonly used in medical imaging to assess perceptual similarity between synthetic and original CT slices; and the Peak Signal-to-Noise Ratio (PSNR), which evaluates reconstruction fidelity. We further introduce the concept of a Visual Turing Test (VTT), wherein expert radiologists attempt to distinguish synthetic from real images as a subjective but clinically pertinent validation. Finally, we reference recent studies—such as Gugulothu et al. (2023)—that employ GANs for both image synthesis and enhancement, demonstrating improvements in downstream classification performance when augmented with high-quality synthetic data.

In summary, GANs provide innovative solutions for pulmonary nodule analysis by generating realistic synthetic data that mitigates the challenges of data scarcity and class imbalance. Studies by Sengodan et al. \cite{sengodan2023early}, Ling et al. \cite{zhu2021hr}, Gugulothu et al. \cite{gugulothu2023novel}, Zhu et al. \cite{zhu2021functional}, and Tyagi et al. \cite{tyagi2022cse} demonstrate the effectiveness of GAN-based augmentation in enhancing the performance and robustness of detection and segmentation models. These findings underscore the growing potential of GANs in clinical applications for pulmonary nodule detection, presenting a promising avenue for improving diagnostic accuracy and reliability. As GAN technology continues to evolve, its remarkable ability to generate high-quality synthetic data has attracted increasing attention, highlighting its critical role in the development of next-generation CAD systems for early lung cancer detection and diagnosis.

\subsection{Transformer Based Methods}
Transformers, first introduced by Vaswani et al. \cite{vaswani2017attention} in 2017, have significantly advanced deep learning, particularly through the adoption of self-attention mechanisms \cite{ji2024sine}. Unlike traditional models that process input data sequentially or focus on local features, Transformers can capture long-range dependencies and relationships across the entire input sequence, which is crucial for tasks like medical imaging. The self-attention mechanism, a core component of the Transformer, operates by calculating relationships between different parts of the input using three key matrices: query, key, and value. This allows the network to focus on the most relevant features, enabling it to learn complex patterns and improve performance in tasks such as image classification \cite{zhang2024jointvit,wu2024xlip}, segmentation \cite{tan2024segstitch}, and detection. The basic structure is shown in {\hyperref[fig:7]{Fig. \ref*{fig:7}}}b.

Building on this architecture, the Vision Transformer (ViT) adapts the Transformer model for image analysis by dividing images into patches and treating them as a sequence. By leveraging self-attention, ViT effectively captures both global and local image features, which is particularly beneficial in medical applications like pulmonary nodule detection in CT scans. Recent studies, such as those by Hui Zhang et al. \cite{zhang2022pulmonary} and Chi et al. \cite{chi2024lgdnet}, have demonstrated that incorporating Transformer-based models significantly improves detection sensitivity and reduces false positives. Chi et al. showed that integrating Transformer layers into pulmonary nodule classification tasks resulted in higher accuracy and better interpretability. Similarly, Han Yang et al. \cite{yang2023lung} explored the application of uncertainty-aware attention in UGMCS-Net, further enhancing segmentation performance, particularly in complex nodule shapes and low-confidence regions.

In addition to ViT, hybrid models like the Swin Transformer have been developed to handle high-resolution medical images by introducing a hierarchical structure. This design allows the model to process images at multiple scales, capturing detailed features across different levels of granularity. For instance, Jinjiang Liu et al. \cite{liu2024multiscale} incorporated 3D coordinate attention and edge enhancement in SCA-VNet, achieving a high Dice coefficient by improving edge detection in pulmonary nodule segmentation. Similarly, Chenglong Wang et al. \cite{wang2024towards} utilized attention gating and multi-task learning in ExPN-Net, improving both segmentation and classification with an AUC of 0.992. These Transformer-based approaches, especially the Vision Transformer and its variants, have demonstrated significant potential in advancing the accuracy, sensitivity, and efficiency of early lung cancer detection and diagnosis.

\subsection{Hybrid Deep Learning Methods}
Hybrid deep learning methods combine different types of neural networks into a unified framework to tackle complex tasks, such as pulmonary nodule detection, segmentation, and classification. By leveraging the unique strengths of various models, including CNNs, RNNs, GANs, and attention mechanisms, these hybrid methods aim to enhance overall performance, stability, and robustness. The primary advantage of hybrid models is their ability to address the limitations of individual networks, providing better feature extraction, improving accuracy, reducing false positives, and optimizing computational efficiency, particularly in challenging tasks like medical image analysis.

\begin{figure*}
  \begin{minipage}[t]{\linewidth}
    \centering
    \includegraphics[width=1\textwidth]{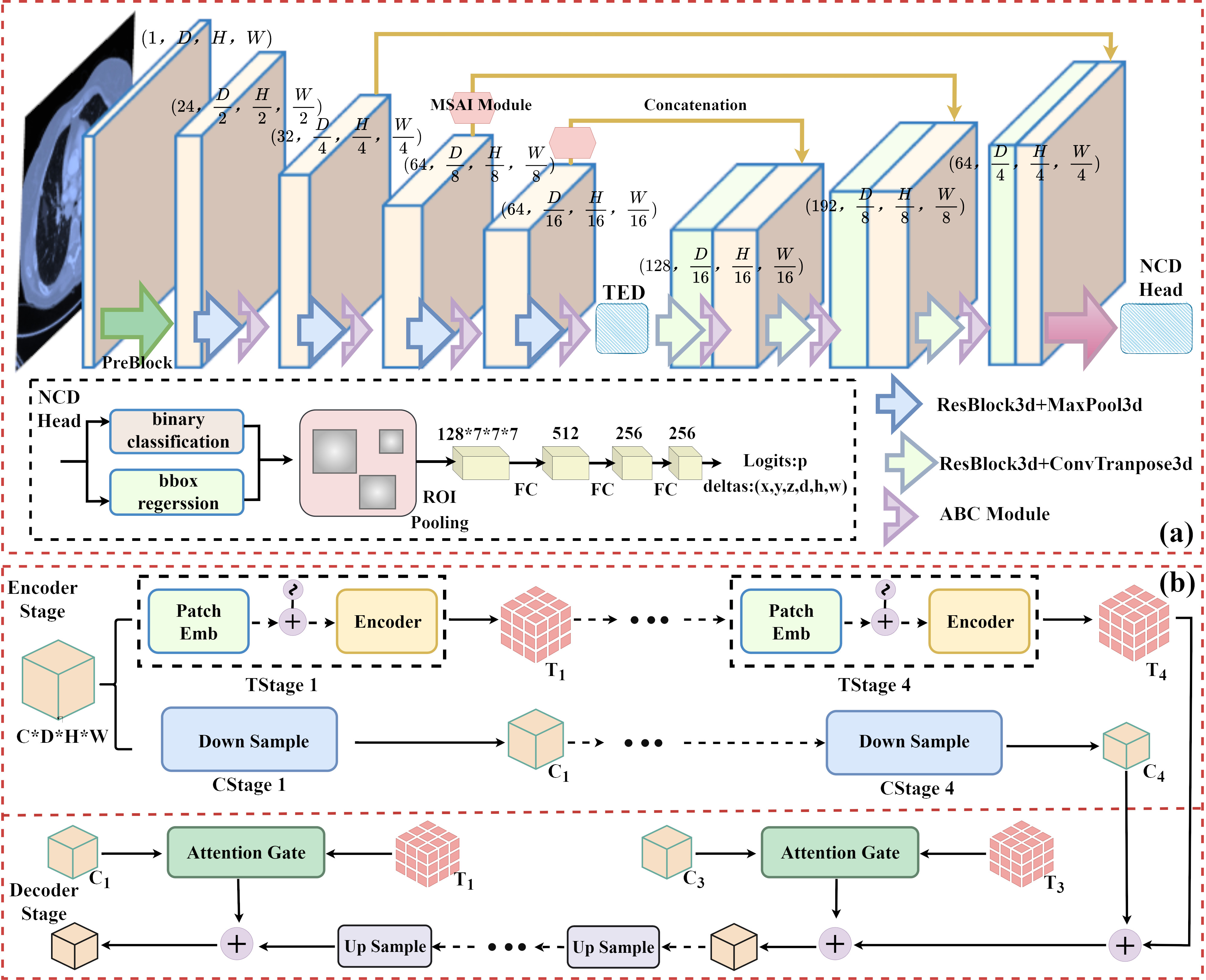}
\caption{Some structural examples of hybrid deep learning models: (a) Combination of 3D CNN, ACB, MSAI, TED, and dual heads for NCD and FPR \cite{ma2024ticnet}, which enhances multi-scale feature extraction and improves both classification and localization in nodule detection. (b) Combination of CNN and Transformer branches with attention gates for local and global feature coupling \cite{chi2024lgdnet}, allowing efficient integration of local and global features and guiding the decoding process for better feature reconstruction and detection accuracy in medical images.}
    \label{fig:8}
  \end{minipage}
\end{figure*}

For example, Rashid et al. \cite{rashid2024nodule} developed a framework combining local binary patterns (LBP) with Histogram of Oriented Surface Normal Vectors (HOSNV) features, achieving a sensitivity of 98.49\% in detecting nodules. Cai et al. \cite{cai2024msdet} proposed MSDet, a model that combines ERD, PCAM, and TODB to capture richer contextual information and reduce false positives caused by nodule occlusion through an extended receptive domain strategy, improving small pulmonary nodule detection and achieving an 8.8\% mAP improvement on the LUNA16 dataset. Siqi Liu et al. \cite{liu2020no} enhanced detection robustness by incorporating adversarial attacks into CNNs, improving the system's ability to detect rare nodules under noisy conditions. Similarly, Safta et al. \cite{safta2024advancing} integrated 3D-LOP descriptors with 3D-CNN features, achieving an accuracy of 97.84\% and an AUC of 0.9912. As shown in {\hyperref[fig:8]{Fig. \ref*{fig:8}}}a, Ling Ma et al. \cite{ma2024ticnet} introduced TiCNet, combining Transformer and CNN architectures with multi-scale feature fusion, which achieved a CPM of 90.73\% and a sensitivity exceeding 93\%, demonstrating its efficacy in reducing false positives. These hybrid approaches show that combining diverse feature extraction and classification techniques significantly improves pulmonary nodule detection and classification accuracy.

Other studies further highlight the versatility of hybrid deep learning methods. As shown in {\hyperref[fig:8]{Fig. \ref*{fig:8}}}b, Jianning Chi et al. \cite{chi2024lgdnet} and Gonidakis et al. \cite{gonidakis2023handcrafted} demonstrated the effectiveness of combining local and global feature representations using attention gates and hand-crafted features, respectively. Gonidakis et al.'s method notably reduced the required training data by up to 43\% while maintaining high performance. Additionally, Mao et al. \cite{mao2021hessian} employed a Hessian-MRLoG method to enhance contrast and reduce false positives, achieving a detection accuracy of 93.6\%. These strategies illustrate the benefits of integrating various deep learning techniques to enhance segmentation and detection performance.

The successful applications of hybrid models, such as those proposed by Gugulothu et al. \cite{gugulothu2023novel}, Usman et al. \cite{usman2024meds}, and Pinheiro et al. \cite{de2020detection}, highlight the effectiveness of integrating advanced algorithms to improve detection accuracy, reduce false positives, and enhance efficiency in lung cancer analysis. However, these hybrid models also introduce challenges. Their complexity can lead to overfitting, requiring additional regularization or improved generalization strategies. Furthermore, the combination of multiple components increases computational demands, leading to higher costs and longer processing times. Effectively integrating different architectures and performing hyperparameter tuning for such complex systems further adds to the difficulty. To address these challenges, more efficient model architectures, advanced optimization techniques, larger datasets, and automated hyperparameter tuning methods could be explored to improve performance and scalability.

Recently, Mamba models \cite{gu2023mamba, dao2024transformers} have gained prominence for their ability to effectively capture long-range dependencies \cite{zhang2024motion,zhang2024infinimotion}, significantly enhancing performance in dense prediction tasks like semantic segmentation. Studies by Bian et al. \cite{bian2024mambaclinix} and Zhang et al. \cite{zhang2024vm} have demonstrated the significant potential of the Mamba architecture in medical image segmentation through the development of the MambaClinix model and VM-UNetV2 framework. MambaClinix combines the Mamba architecture with Hierarchical Gated Convolutional Networks (HGCN), achieving an effective balance between local and global features in 3D medical images. Its lower-level modules capture high-order spatial relationships, while the higher-level modules utilize the Mamba architecture to extract long-range dependencies, resulting in remarkable segmentation performance in datasets involving lung and liver tumors. Simultaneously, VM-UNetV2 introduces a Vision State Space (VSS) model and a Semantics and Detail Infusion (SDI) module, using linear computational complexity to efficiently capture contextual information. Compared to traditional CNNs and Transformer-based models, VM-UNetV2 has demonstrated superior performance in medical segmentation tasks. This trend represents a significant advancement in intelligent medicine, particularly in medical image analysis, where hybrid models incorporating the Mamba architecture \cite{gu2023mamba, dao2024transformers} are overcoming the limitations of traditional methods in terms of computational efficiency and global feature extraction. As the volume of medical data continues to grow, models capable of efficiently modeling long-range dependencies and handling complex contextual information will become increasingly essential. These advancements not only enhance early disease detection and diagnosis but also lay a strong foundation for the widespread adoption of intelligent medical systems.

In summary, hybrid deep learning models represent a rapidly growing trend, particularly in the field of pulmonary nodule detection and classification, where these models are showing significant potential. With ongoing advancements in deep learning technologies and the expansion of interdisciplinary research, even more innovative hybrid models are expected to emerge. These models will likely address increasingly complex challenges, driving further progress in the early detection and accurate diagnosis of lung cancer, ultimately improving clinical outcomes.

\section{Literature Search and Screening}
\begin{figure*}
  \begin{minipage}[t]{\linewidth}
    \centering
    \includegraphics[width=1\textwidth]{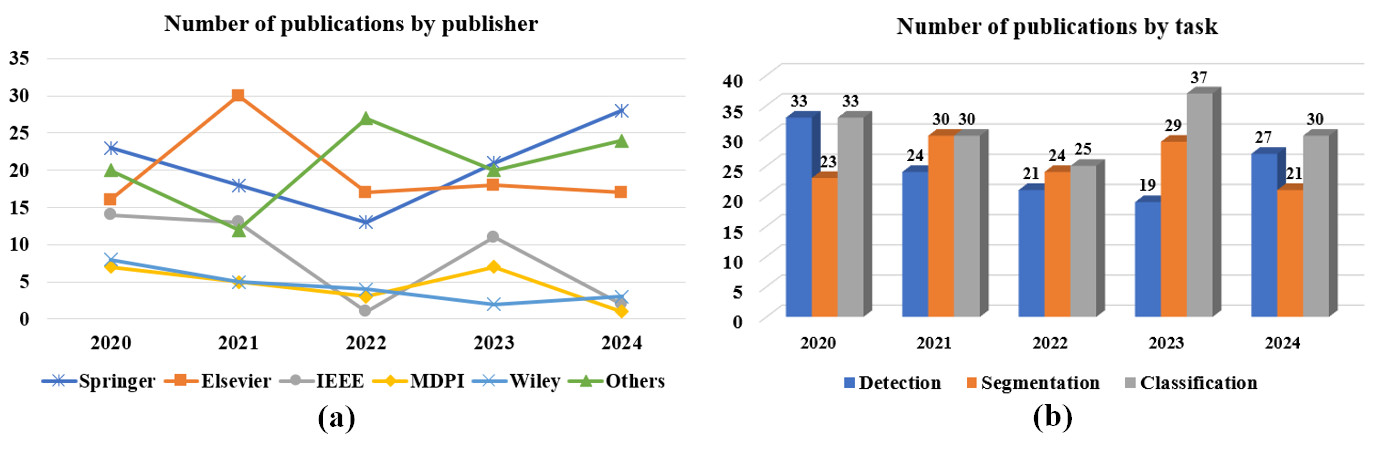}
    \caption{(a) Number of publications in each database/platform from 2020 to 2024. (b) Number of publications on different pulmonary nodule treatment tasks from 2020 to 2024.}
    \label{fig:9}
  \end{minipage}
\end{figure*}

This review systematically retrieved relevant literature from multiple databases, including IEEE Xplore, Wiley Online Library, Google Scholar, Web of Science, ACM, Science Direct, Springer Link, and Engineering Village, covering the period from January 2020 to May 2024. The distribution of publications across these databases is shown in {\hyperref[fig:9]{Fig. \ref*{fig:9}}}a. Through rigorous screening, this review consolidates the applications of machine learning and deep learning techniques in computer-aided diagnosis (CAD) of pulmonary nodules, with a particular focus on their implementation in CT imaging. In addition to highlighting major research achievements, this review discusses the advantages and limitations of current technologies and outlines future research directions, providing valuable insights for researchers in the field.

\begin{figure*}
  \begin{minipage}[t]{\linewidth}
    \centering
    \includegraphics[width=1\textwidth]{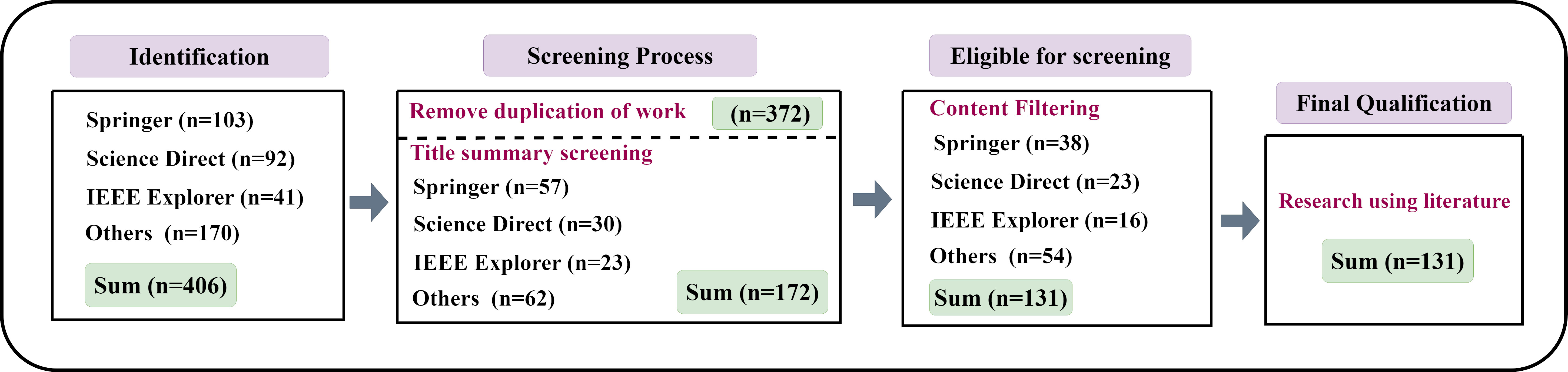}
    \caption{The literature screening process adapted in this study. Screening criteria for step 2 include the following: time range (2020.1 - 2024.5), journal publication, journal type, and language (English); screening criteria for step 3 include the exclusion of methods that do not meet the criteria and conference papers. 'Others' includes publications from Wiley, MDPI, Taylor and Francis, and Hindawi.}
    \label{fig:10}
  \end{minipage}
\end{figure*}

During the retrieval process, 406 relevant articles were identified, covering different aspects of pulmonary nodule analysis: 124 on detection, 155 on segmentation, and 127 on classification, as illustrated in {\hyperref[fig:9]{Fig. \ref*{fig:9}}}b. After applying stringent screening criteria, 131 articles were selected, with the selection process depicted in {\hyperref[fig:10]{Fig. \ref*{fig:10}}}. The final selection includes 56 articles on detection, 33 on segmentation, and 42 on classification. The subsequent sections will discuss the technical approaches, research outcomes, and applications derived from these studies, providing a comprehensive overview of the CAD landscape for pulmonary nodules and offering a thorough understanding of the current research status and future trends in this evolving field.

\section{Datasets and Benchmarks} 
The development of Computer-Aided Diagnosis (CAD) systems for pulmonary nodules relies heavily on high-quality imaging data and robust evaluation metrics. Publicly available datasets serve as foundational resources for training, testing, and validating deep learning algorithms, facilitating standardized comparisons across different methods. These datasets vary in size, annotation quality, and imaging techniques, making them critical for different stages of research and clinical applications. In addition to datasets, benchmarks play a key role in evaluating the performance of CAD systems, providing objective metrics such as accuracy, sensitivity, and specificity to assess algorithm effectiveness. This section provides an overview of the key datasets widely used in CAD research for pulmonary nodules and discusses the most commonly employed benchmarks and evaluation metrics. As shown in Table ~\ref{table:2}, these datasets provide essential resources for advancing research in pulmonary nodule detection and diagnosis.

The LIDC-IDRI dataset \cite{armato2011lung} is one of the most comprehensive publicly available lung CT datasets, jointly created by the National Cancer Institute (NCI) and the American College of Radiology (ACR). It contains over 1,000 lung CT cases, annotated by four experienced thoracic radiologists through a multi-phase, consensus-based evaluation. Each scan is annotated with regions of interest, specifically focusing on pulmonary nodules of varying sizes, and includes both nodule and non-nodule cases. The diversity of imaging protocols and scanning devices makes this dataset widely applicable for CAD system development. Furthermore, its comprehensive annotations from different radiologists allow for the exploration of inter-observer variability, a key challenge in medical image analysis.

The LUNA16 dataset \cite{setio2017validation} is derived from LIDC-IDRI but focuses specifically on pulmonary nodule detection. It includes 888 high-quality CT scans with detailed annotations of nodules larger than 3 mm in diameter. LUNA16 also provides a standardized evaluation framework, making it a widely adopted benchmark for comparing nodule detection algorithms. The dataset includes annotations that highlight the location and size of each nodule, offering a valuable resource for both segmentation and detection tasks. The nodule cases are categorized based on their risk levels, helping researchers more effectively distinguish between benign and malignant nodules.

The ELCAP dataset \cite{henschke2001early} is part of the Early Lung Cancer Action Project and consists of low-dose CT (LDCT) scans, primarily targeting high-risk populations, such as heavy smokers. ELCAP emphasizes early-stage lung cancer screening and includes a large number of small nodules, making it an essential resource for early detection studies. The dataset includes detailed clinical and demographic information, such as patient age, smoking history, and risk factors. Although only part of this dataset is publicly available, its focus on high-risk groups makes it valuable for developing screening algorithms aimed at detecting lung cancer in its earliest stages.

The NSCLC-Radiomics dataset \cite{aerts2014decoding} focuses on non-small cell lung cancer (NSCLC) and includes not only CT scans but also extensive clinical and molecular data. This dataset contains over 400 CT scans from patients with NSCLC, annotated with tumor regions, genomic information, and treatment outcomes. Its richness in clinical context makes it particularly valuable for research into personalized treatment strategies and prognostic modeling. The dataset also supports the study of radiomics, which involves extracting a large number of features from medical images to uncover disease characteristics that are not visible to the naked eye, providing a bridge between medical imaging and molecular biology.

\begin{table}[h!]
\centering
\small
\caption{Overview of Datasets Commonly Adapted in Pulmonary Nodule Computer-Aided Diagnosis (CAD) Research.}
\label{table:2}
\begin{tabular}{lcccc}
\toprule
\textbf{Dataset}      & \textbf{No. of Scans} & \textbf{Focus Area}        & \textbf{Use Case} & \textbf{Annotations} \\
\midrule
LIDC-IDRI             & 1,018                 & Pulmonary nodule Detection      & Detection, Classification, Segmentation & \checkmark \\
LUNA16                & 888                   & Nodule Detection (Nodules >3mm)  & Detection Benchmark & \checkmark \\
ELCAP                 & 50                & Early Lung Cancer Screening & Early Detection & \checkmark \\
NSCLC       & 422                   & Non-Small Cell Lung Cancer  & Personalized Treatment, Prognosis & \checkmark \\
ANODE09               & 55                    & Nodule Detection (Various types) & Benchmarking Detection Algorithms & \checkmark \\
PN9               & 8798                    & Nodule Detection \& Classification & Large-scale Model Training \& Validation & \checkmark \\
\bottomrule
\end{tabular}
\end{table}

The ANODE09 dataset \cite{van2010comparing} was developed for an international competition on automatic pulmonary nodule detection and consists of 55 CT scans that cover a wide range of nodule types, sizes, and shapes. This dataset focuses on providing a controlled environment for evaluating detection algorithms. The standardized evaluation metrics, such as sensitivity and false positive rates, facilitate direct comparisons between different methods. Although smaller in size compared to other datasets, ANODE09 is highly structured and provides clear, well-defined cases that are ideal for benchmarking new detection techniques. The diversity of nodule types, including solid, subsolid, and ground-glass opacities, presents an additional challenge for algorithm development, making this dataset particularly useful for testing algorithm robustness.

The PN9 dataset \cite{mei2021sanet}, released in 2020, is a relatively new and valuable resource for pulmonary nodule analysis. It contains high-resolution CT images collected from multiple institutions and scanners, enhancing its diversity and representativeness. One of its key strengths lies in the inclusion of subtle and diagnostically challenging nodules, which are often underrepresented in older datasets. Additionally, PN9 offers clinical annotations that are closely aligned with contemporary diagnostic standards, making it particularly suitable for training and validating models intended for real-world clinical deployment.

In terms of evaluating the effectiveness of CAD systems, several standardized benchmarks and evaluation metrics are employed. One widely adopted benchmark for nodule detection is the Competition Performance Metric (CPM), based on the Free-Response Receiver Operating Characteristic (FROC) curve. CPM evaluates an algorithm’s sensitivity at predefined false positive rates, typically ranging from 1/8 to 8 false positives per scan. By averaging sensitivity at these rates, CPM provides a robust measure of how well a model balances sensitivity and false positives. The FROC curve, which plots the true positive rate (sensitivity) against the average number of false positives per scan, is a key tool in CAD research. The area under the FROC curve (AUC) indicates overall performance, with higher AUC values (ranging from 0.5 to 1) reflecting better discrimination between positive and negative cases.

For segmentation tasks, the Dice Similarity Coefficient (DSC) is commonly employed to assess how accurately an algorithm delineates nodule boundaries, with higher DSC values indicating better overlap between predicted and actual segmentations. In classification tasks, key metrics include sensitivity (the ability to correctly identify nodules), specificity (the ability to correctly identify non-nodules), and accuracy, which measures the overall correctness of the model’s predictions. These metrics—sensitivity, AUC, specificity, and DSC—are critical for balancing nodule detection and minimizing false positives, ultimately guiding the development of reliable CAD systems for clinical practice.

We also conducted a detailed analysis of distributional imbalances in the two most widely used public datasets, LIDC-IDRI and LUNA16. In LIDC-IDRI, benign nodules account for 82\% (3,572 / 4,356) of all annotated nodules, leading to a strong class skew. Gautam et al. (2024) have shown that models trained on this imbalanced distribution suffer a 12.8\% drop in sensitivity when evaluated on clinical cohorts with a more realistic 45\% benign‐to‐malignant ratio. Moreover, both datasets exhibit a size imbalance, with the vast majority of nodules falling in the 3–10 mm range and very few examples of very small or large nodules; a label imbalance, where certain subtypes (e.g., part-solid or ground-glass) are underrepresented; and an anatomical imbalance, as nodules cluster in specific lung regions. To help mitigate some of these issues, we have updated Table ~\ref{table:2} to include the more recently released PN9 (2020) dataset, which provides a larger and more evenly distributed sample across nodule sizes, subtypes, and acquisition sources.

\section{Discussion: Key Techniques and Breakthroughs in Lung Cancer Detection, Segmentation, and Classification}
In recent years, AI-driven advancements in computer-aided diagnosis (CAD) systems for pulmonary nodules have transformed the field of lung cancer detection, segmentation, and classification. These innovations, bolstered by high-quality datasets such as LIDC-IDRI and LUNA16, have not only accelerated technological progress but also fostered international research collaboration. In detection, the integration of deep learning models, especially multi-task convolutional neural networks, has significantly enhanced the sensitivity and accuracy of identifying nodules from CT scans. In segmentation, advanced models like UNet, 3D CNNs, and GANs have greatly improved precision in delineating nodule boundaries, a critical factor in evaluating nodule size and growth. For classification, the combination of deep learning techniques and texture-based features has refined the differentiation between benign and malignant nodules, thus improving diagnostic reliability. Together, these innovations have optimized CAD systems, paving the way for broader clinical applications and improved patient outcomes through earlier and more accurate lung cancer diagnoses. The following sections will examine recent progress in detection, segmentation, and classification, highlighting the key techniques and breakthroughs in each area and providing a critical analysis of their clinical relevance and future directions.

\subsection{Discussion on AI-Driven Advances in Lung Cancer Detection}
Detecting pulmonary nodules is essential for early lung cancer diagnosis and treatment, enabling timely intervention. Pulmonary nodule detection generally involves two stages: screening potential nodules in CT scans and classifying malignancy likelihood, which guides clinical decisions. Recent advancements in AI, particularly deep learning models like multi-task CNNs, YOLOv4, UNet, and GANs, have significantly improved detection sensitivity and accuracy, enhancing CAD systems. High-quality datasets such as LIDC-IDRI and LUNA16 have facilitated these advances by providing annotated data for effective model training. Table ~\ref{table:3} summarizes 19 recent studies (2020-2024) utilizing the LIDC-IDRI dataset, highlighting their strengths, limitations, and clinical relevance.

\begin{table*}
\scriptsize %
\setlength{\tabcolsep}{0.13cm} %
\renewcommand{\arraystretch}{1.7} %
\setlength{\heavyrulewidth}{1pt} %
    \centering 
   \caption{Pulmonary nodule detection methods and their performance in CT images from the LIDC-IDRI dataset.}
\label{table:3}
\begin{tabular}{@{\extracolsep{\fill}} p{1.5cm} p{0.7cm} p{0.7cm} p{1.6cm} p{1.6cm} p{1.6cm} p{1.6cm} p{1.6cm} p{1.4cm} | p{1.8cm} @{}}
\toprule
  \makecell[l]{Authors} & \makecell[l]{Year} & \makecell[l]{2D/3D} & \makecell[l]{Technique/\\Network Type} & \makecell[l]{Training/\\Validation/\\Testing Set} & \makecell[l]{Software/\\Hardware\\Utilized} & \makecell[l]{Main\\ Method} & \makecell[l]{Nodules\\ Types} & \makecell[l]{Validation/\\Test Method} & \makecell[l]{Model\\Performance} \\
  \midrule
  \specialrule{0.1pt}{0pt}{0pt} 
  Veronica et al. {\cite{veronica2020effective}} & 2020 & 2D and 3D & ANN, Fuzzy C-Means & LIDC-IDRI, ELCAP & MATLAB, 4GB RAM & FCM,ANN with OALO & Benign, malignant &  & Sensitivity: 96.10\% \\
  Bhatt et al. {\cite{bhatt2022diagnosis}} & 2022 & 2D & YOLOv4 & Tra 70\%, Val 20\%, Test 10\% & TensorFlow RTX 2070 & YOLOv4 model & Various nodule types and sizes &  & YOLOv4: 81.00\% \\
  Harsono et al. {\cite{harsono2022lung}} & 2022 & 3D & I3DR-Net & LIDC: 605 tra, 202 val, 202 tes & Pytorch, Tesla P100 16GB & Inflated 3D ConvNet (I3D) & Solid and non-solid nodules &  & Sensitivity: 94.60\% \\
  Nair et al. {\cite{nair2022prediction}} & 2022 & 2D & RWI segmentation, RF, KNN & 534 training, 150 testing & MATLAB 2018a & Random Walker with Random Forest, k-NN & Benign and malignant nodules &  & RWI-RF: 99.99\%; RWI-k-NN: 98.9919\% \\
  Jain et al. {\cite{jain2023pulmonary}} & 2023 & 3D & Two-stage CNN, U-net & LIDC-IDRI: 5023tra/val & Python, Tesla P100 & Two-stage CNN, U-net & Pulmonary nodules $\geq$ 3 mm & Train-Validation-Test Split & Sensitivity: 83.55\% \\
  Gugulothu et al. {\cite{gugulothu2023novel}} & 2023 & 2D & CSDR-J-WHGAN & Training 80\%, Testing 20\% & Python & SDMMT, MD-HHOA, CSDR-J-WHGAN & Nodule and non-nodule &  & Sensitivity: 96.98\% \\
  Rashid et al. {\cite{rashid2024nodule}} & 2024 & 2D and 3D & HOSNV, LBP, XGBoost & LIDC-IDRI & - & XGBoost & Vessel, isolated, juxtapleural &  & XGBoost: 98.49\% \\
  Agnes et al. {\cite{agnes2024wavelet}} & 2024 & 2D & Wavelet U-Net++ & LIDC-IDRI: 1018 scans & Python Keras, GPU & U-Net++, hybrid loss & Small, irregular nodules &  & DSC: 93.70\%; IoU: 87.80\% \\  
  Gautam et al. {\cite{gautam2024lung}} & 2024 & 2D & ResNet-152, EfficientNet-B7 & LIDC-IDRI & - & Weighted average ensemble & Benign and malignant &  & Sensitivity: 98.60\% \\
  Gugulothu et al. {\cite{gugulothu2024early}} & 2024 & 2D & Hybrid DL (HDE-NN) & LIDC-IDRI & - & Hybrid differential evolution-based NN & Benign and malignant &  & Sensitivity: 95.25\% \\
  \midrule
  Halder et al. {\cite{halder2020adaptive}} & 2020 & 2D & Segmentation (AMST) & LIDC/IDRI, private dataset & - & Morphological filter with SVM & Solid, subsolid, GGO &  & Sensitivity: 94.88\% \\
  Masud et al. {\cite{masud2020light}} & 2020 & 2D & CNN & LIDC 1279 images & Nvidia GTX1050,12GB & Light CNN & Normal, Benign, malignant &  & Accuracy: 97.90\% \\
  Farhangia et al. {\cite{farhangi2021automatic}} & 2021 & 2D and 3D & Unified CNN & 888 CT exams & CT volume: 12 min & Dilated 1D convolutions & Solid, subsolid, large nodules & 10-fold cross-validation & Sensitivity: >96.00\% \\
  Zheng et al. {\cite{zheng2021deep}} & 2021 & 3D & U-net++, Efficient-Net & LIDC:888scans, 1186 nodules & - & Multiscale dense CNNs & Small nodules (<6 mm) &  & Sensitivity: 94.20\% \\
  Chen et al. {\cite{chen2023novel}} & 2022 & 3D & F-Net, MSS-Net & Training 70\%, Validation 20\%, Test 10\% & - & F-Net, MSS-Net & Solitary, varying sizes (3-30 mm) &  & 0.971 (1 FP/scan); 0.978 (2 FPs/scan) \\ 
  Usman et al. {\cite{usman2024meds}} & 2024 & 3D & Multi-encoder, Self-distillation & Training 80\%, Validation 10\% & Python & 3Dsub-volumes, self-distillation & Nodules $\geq$ 3 mm &  & 20.27 Fps/scan \\
  \midrule
  Fu et al. {\cite{fu2021fusion}} & 2021 & 3D & Multi-DL, MLP & 3:1 training/validation & Keras, GTX1080ti & Fusion of 3D lung CT images & Squamous cell carcinoma &  & Specificity: 90.60\% \\  
  Majidpourkhoei et al. {\cite{majidpourkhoei2021novel}} & 2021 & 3D & CNN & LIDC,1415 test images & Python, 8GB RAM & CNN with 3D convolutions & Nodule sizes (1-5 mm) & 5-fold cross-validation & Specificity: 91.70\% \\
  Hung et al. {\cite{hung2023interpretable}} & 2023 & 3D & HSNet & Original: 4252 images, split 3:1 & TensorFlow 2.7, NVIDIA RTX 3060, 32GB & 3D hierarchical semantic CNN (HSNet) & Subtlety, texture, sphericity, malignancy &  & Cal: 98.73\%; Margin: 92.07\%; Subtlety: 90.26\% \\
  \bottomrule
\end{tabular}
\end{table*}

Key Innovations and Performance Comparisons: Over the past several years, significant advancements have been made in AI-driven pulmonary nodule detection, contributing uniquely to the evolution of the field. In 2020, Masud et al. \cite{masud2020light} developed a lightweight deep learning model achieving 97.9\% accuracy, showcasing the potential for AI deployment in resource-constrained environments, such as mobile devices in clinics with limited computational power. In 2022, Bhatt et al. \cite{bhatt2022diagnosis} demonstrated the utility of YOLOv4 for real-time detection with 95\% precision, emphasizing the practical application of AI in rapid diagnostics. Fu et al. \cite{fu2021fusion} integrated 3D CT imaging with serum biomarkers, highlighting a multimodal approach that combines diverse data sources to enhance diagnostic precision. In 2023, Gugulothu et al. \cite{gugulothu2023novel} introduced a hybrid deep learning approach, and by 2024, Gautam et al. \cite{gautam2024lung} combined ResNet, DenseNet, and EfficientNet to achieve a sensitivity of 98.6\%, significantly reducing false negatives. Furthermore, Usman et al. \cite{usman2024meds} developed MEDS-Net, which achieved a CPM score of 93.6\% by reducing false positives, thus minimizing unnecessary invasive procedures and enhancing diagnostic reliability. These innovations collectively advance the clinical applicability of AI for lung cancer detection by improving both accuracy and efficiency. To address dataset imbalances, recent studies have adopted dynamic weighted loss functions and carefully tuned oversampling strategies. Dynamic weighting—where training losses for minority classes are up-weighted on the fly—has been shown to improve recall on underrepresented nodules, reducing false negatives without destabilizing overall training. However, overly aggressive oversampling in LUNA16 can introduce synthetic bias and lead to overfitting, ultimately harming model generalization on unseen data. We therefore highlight a balanced approach: combining moderate oversampling with cost-sensitive learning and regularization techniques to narrow the generalization gap and mitigate diagnostic blind spots in real-world clinical deployments.

Model Adoption and Clinical Implications: The discussed studies highlight a range of technical approaches focusing on improving sensitivity, computational efficiency, and real-time applicability. Models such as YOLOv4 and MEDS-Net illustrate the potential of AI to be integrated into practical, real-time clinical workflows, whereas the multimodal diagnostic model by Fu et al. \cite{fu2021fusion} underscores the value of combining multiple data sources for a more comprehensive diagnostic approach. A crucial factor in clinical adoption is the computational complexity of these models. While high accuracy is desirable, the scalability of complex models like Mask R-CNN and GANs is often limited due to their heavy computational requirements. For instance, Zhu et al.'s \cite{zhu2021functional} GAN-based approach effectively addresses data imbalance but at the cost of significant computational demands, potentially hindering scalability in less resource-rich environments. In contrast, Wu et al.'s \cite{wu2024yolo} enhancement of YOLOv7 presents a more efficient model that balances performance and computational load, making it more practical for broader clinical implementation.

Comparisons Across Benchmark Datasets: Compared to the LIDC-IDRI dataset, LUNA16 employs CPM (Competition Performance Metric) as a benchmark, providing a standardized measure for comparing CAD system performance. This allows for a clearer evaluation of different models' strengths and applicability in various clinical scenarios. Table ~\ref{table:4} compiles key metrics such as author, year, technique type, validation method, and CPM score, offering a comprehensive view of advancements between 2020 and 2024. Studies employing deep learning techniques that integrate 3D contextual information and multi-scale feature fusion, such as Mask R-CNN and 3D CPM-Net, exhibit outstanding performance, improving detection accuracy while minimizing false positives.

\begin{table*}
\scriptsize %
\setlength{\tabcolsep}{0.13cm} %
\renewcommand{\arraystretch}{1.6} %
\setlength{\heavyrulewidth}{1pt}
    \centering 
   \caption{Pulmonary nodule detection methods and their performance in CT images from the LUNA16 dataset.}
\label{table:4}
\begin{tabular}{@{\extracolsep{\fill}} p{1.2cm} p{1.1cm} p{0.7cm} p{1.6cm} p{1.4cm} p{1.6cm} p{1.6cm} p{1.2cm} p{1.4cm} 
 | p{2.2cm} @{}}
\toprule
  \makecell[l]{Authors} & \makecell[l]{Year} & \makecell[l]{2D/3D} & \makecell[l]{Technique/\\Network Type} & \makecell[l]{Training/\\Validation/\\Testing Set} & \makecell[l]{Software/\\Hardware\\Utilized} & \makecell[l]{Main\\Method} & \makecell[l]{Nodules\\Types} & \makecell[l]{Validation/\\Test Method} & \makecell[l]{Model\\Performance} \\
  \midrule
  \specialrule{0.1pt}{0pt}{0pt}  
  Cai et al. {\cite{cai2020mask}} & 2020 & 3D & Mask R-CNN & LUNA16, Ali TianChi & Quadro P6000 GPU & Mask R-CNN, ResNet50, FPN, RPN & Benign and malignant &  & 88.1\% at 1 FP/scan; 88.7\% at 4 FP/scan ; AP@50: 88.2\% \\
  Song et al. {\cite{song2020cpm}} & 2020 & 3D & 3D CPM-Net & LUNA16 (888 scans) & SenseCare Platform & Anchor-free, 3D context & Benign and malignant &  & 91.2\% at 1 FP/scan; 92.4\% at 2 FP/scan \\
  Yuan et al. {\cite{yuan2021efficient}} & 2021 & 3D & Multi-path 3D CNN & LUNA16, split 8:1:1 & TensorFlow, RTX2080 TI GPU & Multi-path 3D CNN, feature fusion & Various sizes &  & 0.952 at 4 FP/scan; 0.962 at 8 FP/scan ; CPM: 0.881 \\
  Agnes et al. {\cite{agnes2022two}} & 2022 & 2D and 3D & Two-stage Pulmonary nodule detection & 23,720 CT, 80\% Train, 20\% Test & Keras, GTX 1060 (6 GB) & Atrous UNet+, PD-CLSTM & Small nodules(5-9 mm) &  & 0.544-0.986 at different FP levels ; CPM: 0.93 \\
  Zhao et al. {\cite{zhao2022pulmonary}} & 2022 & 2D and 3D & Faster R-CNN & LUNA16 & Platform - Colab; Ubuntu; K80 GPU & Multiscale fusion & Small nodules & Train-Validation-Test Split & 0.905 at 4 FP/scan ; CPM: 0.829 \\
  Gonidakis et al. {\cite{gonidakis2023handcrafted}} & 2023 & 3D & Handcrafted features + CNN & 70\% Train, 10\% Val, 20\% Test & Keras with TensorFlow, 2080 RX & Handcrafted + CNN & Benign and malignant &  & 0.5 mean FP/scan ; Fusion: 94.3\% \\
  Hendrix et al. {\cite{hendrix2023deep}} & 2023 & 2D and 3D & AI system & 500 CT (Hospital A), 888 (LUNA16) & - & AI, multi-view ResNet50, FROC & Actionable benign nodules &  & Internal: 90.9\%; External: 92.4\%; Benign: 94.3\% \\
  Wu et al. {\cite{wu2024yolo}} & 2024 & 2D & YOLO-MSRF & LUNA16 & - & YOLO-MSRF & Benign and malignant &  & Sensitivity: 94.02\%; mAP: 95.26\% \\
  Zhang et al. {\cite{zhang2024s}} & 2024 & 3D & S-Net, U-shaped & LUNA16 & - & S-Net, hybrid loss & Different shapes &  & S-Net(R): 0.914; S-Net(S): 0.915 \\
  Budati et al. {\cite{budati2024intelligent}} & 2024 & 2D & SbYSF, Sailfish + YOLO & LUNA16 & Python in Windows 10 & Sailfish-based YOLO (SbYSF) & Lung cancer nodules &  & CPM: 99.75\% \\
  \midrule  
  Cao et al. {\cite{cao2020two}} & 2020 & 3D & Two-Stage CNN & LUNA16, 10\% validation & Ubuntu 14.04, Python 3.6.4, GTX-1080Ti & TSCNN: UNet, 3D CNN & Calcific, cavitary &  & DenseNet: 0.768; SeResNet: 0.730; IncepNet: 0.686 \\
  Zhu et al. {\cite{zhu2021functional}} & 2021 & 3D & FRGAN & 543 images, augmented to 0.22M & Adam optimizer, learning rate $1.0 \times 10^{-4}$ & GAN augmentation, 3D-CNN & Solid, subsolid, large nodules &  & CPM: 0.915 \\
  Shi et al. {\cite{shi2021automatic}} & 2021 & 3D & 3D Res-I, Faster R-CNN & Fold 1-9 Train, Fold 0 Test & Ubuntu 16.04, PyTorch, GTX 1080Ti & U-Net-Like network, Faster R-CNN & Solid, semi-solid &  & Sensitivity: 96.37\%; FROC: 83.75\% \\
  Zhang et al. {\cite{zhang2022pulmonary}} & 2022 & 3D & 3D MSA, 3D Faster R-CNN & LUNA16, TianChi & - & Multiscale attention & Various sizes & 10-fold cross-validation & 0.945 at 1 FP/scan ; CPM: 0.927\\
  Zhang et al. {\cite{zhang2023lungseek}} & 2023 & 3D & LungSeek, 3D SK-ResNet & LUNA16: 9 Train subsets (800 CT) & PyTorch, GTX 1080, 16 GB RAM & LungSeek, 3D SK-ResNet & Benign and malignant &  & SK-ResNet: 95.78\%; Res18: 95.53\%; DPN: 94.32\% \\
  Gao et al. {\cite{gao2024robust}} & 2024 & 3D & FULFIL, GCN & LUNA: 140 scans & PyTorch 1.1.0, RTX 3090 & FULFIL, GCN, Teacher–Student & Benign and malignant &  & 0.574 at 0.125 FP/scan \\
  Lin et al. {\cite{lin2024development}} & 2024 & 3D & 3D RPN & LUNA16, LNOP, LNHE & PyTorch, RTX 2070, 24GB RAM & Modified 3D RPN, CSP-ResNeXt, FPN & Different solid &  & 96.6\% at 8 FP/scan for LUNA16 ; CPM: 90.10\% \\  
  \bottomrule
\end{tabular} 
\end{table*}

Future Directions and Clinical Potential: The development and adoption of AI-driven CAD systems for pulmonary nodule detection must prioritize clinical integration and adaptability. Future research should focus on reducing computational complexity to make these models suitable for environments with limited resources, further developing lightweight and real-time applications to expand access, and integrating multimodal data, such as combining imaging with biomarkers, to improve diagnostic precision. In conclusion, advancements in AI-driven pulmonary nodule detection have significantly enhanced detection sensitivity, reduced false positives, and improved clinical applicability, paving the way for broader adoption. Successful integration into clinical practice, with a continued emphasis on balancing efficiency and accuracy, is essential for realizing the full impact of these systems on patient care.

Single-Stage Detectors for Real-Time Nodule Analysis: Single-stage object detectors such as YOLO-v5 and YOLO-v6 have emerged as powerful tools that balance high detection accuracy with real-time inference, making them ideally suited for integration into clinical workflows. YOLO-v5 bypasses the region-proposal stage by predicting bounding boxes and class probabilities in a single network pass; for instance, Ahmadyar et al. (2024) combined a hierarchical YOLO-v5s with a 3D CNN backbone on LUNA16, achieving 97.8 \% sensitivity at a 0.3 confidence threshold \cite{ahmadyar2024hierarchical}. Wu et al. (2024) further refined this framework with their YOLO-MSRF variant—incorporating multi-scale feature refinement—to reach 94.02 \% sensitivity and 95.26 \% mAP on the same benchmark \cite{wu2024yolo}. YOLO-v6 further advances the architecture by employing an optimized backbone and anchor-free detection head to reduce latency and enhance small-object recall—key for identifying subtle or diminutive nodules—and early reports indicate over 98 \% sensitivity on LIDC-IDRI with per-slice inference times under 50 ms. These developments underscore a trend toward embedding lightweight, fast, and accurate CAD tools directly within PACS or imaging consoles, thereby broadening access to AI-driven lung cancer screening in routine clinical practice.

\subsection{Discussion on AI-Driven Advances in Lung Cancer Segmentation}
Pulmonary nodule segmentation is vital for early lung cancer diagnosis, involving the precise identification of nodules in CT images for volumetric analysis, growth rate estimation, and characterization—critical for treatment strategies. Segmentation quality directly impacts clinical decisions, making it essential for CAD systems. Recent AI advancements, particularly in deep learning models like UNet, GANs, and 3D CNNs, have enhanced segmentation accuracy, supported by datasets such as LIDC-IDRI and LUNA16. Table ~\ref{table:5} summarizes recent studies, detailing model performance, datasets, and validation methods, helping identify suitable models for clinical use. Ultimately, improved segmentation accuracy supports early diagnosis, treatment planning, and better patient outcomes.

\begin{table*}
\scriptsize %
\setlength{\tabcolsep}{0.15cm} %
\renewcommand{\arraystretch}{1.2} %
\setlength{\heavyrulewidth}{1pt}
    \centering 
   \caption{Methods and their performance in pulmonary nodule segmentation on CT images.} 
\label{table:5}
\begin{tabular}{@{\extracolsep{\fill}} p{1.55cm} p{1.1cm} p{1.6cm} p{1.4cm} p{2.0cm} p{1.8cm} p{1.4cm} | p{3.4cm} @{}}
\toprule
  \makecell[l]{Authors} & \makecell[l]{Year} & \makecell[l]{Dataset} & \makecell[l]{Technique/\\Network Type} & \makecell[l]{Main\\Method} & \makecell[l]{Software/\\Hardware\\Utilized} & \makecell[l]{Validation/\\Test Method} & \makecell[l]{Model\\Performance} \\
  \midrule
  \specialrule{0.1pt}{0pt}{0pt}  
  Dutande et al. {\cite{dutande2021lncds}} & 2021 & LIDC-IDRI, LNDb, ILCID & 2D-3D cascaded CNN & SquExUNet: Segmentation & Keras, NVIDIA P100 16GB &  & Segmentation: 80.00\% ; Detection: 90.01\% \\  
  Kadia et al. {\cite{kadia2021r}} & 2021 & LUNA16, VESSEL12 & R2U3D & R2U3D with 3D convolution. & Keras, NVIDIA RTX 2080 Ti &  & Soft-DSC: 0.9920 \\
  Liu et al. {\cite{liu2024multiscale}} & 2024 & LIDC-IDRI & SCA-VNet & Residual edge enhancement & PyTorch 1.12.1 &  & DSC: 87.50\% ; Sensitivity: 86.80\% ; Precision: 88.32\% \\
  Agnes et al. {\cite{agnes2024wavelet}} & 2024 & LIDC-IDRI & Wavelet U-Net++ & U-Net++ with wavelet pooling & Python Keras, GPU &  & DSC: 93.7\% ± 0.14 \\
  Cai et al. {\cite{cai2020mask}} & 2020 & LUNA16, TianChi & Mask R-CNN & Mask R-CNN, resnet50 backbone & Quadro p6000 (1.08GB) &  & 88.1\% at 1 false positive/scan; 88.7\% at 4 false positives/scan \\
  Osadebey et al. {\cite{osadebey2021three}} & 2021 & LIDC-IDRI, 3DIRCAD, ILD, PHTM & CNNs and U-net & Preprocessing: CNN classifier; Processing: U-net & MATLAB, Windows 10, Intel i7-8650U & Train-Validation-Test Split & 3DIRCAD: 0.76-0.95; ILD: 0.81-0.95 \\
  Zhu et al. {\cite{zhu2021hr}} & 2021 & LUNA16 & HR-MPF & HR-MPF with PDM & PyTorch, Intel i7-10700, GTX 2070 &  & DSC: 0.9373 ; Sensitivity: 0.9377 ; Precision: 0.9427 \\
  Tyagi et al. {\cite{tyagi2022cse}} & 2022 & LUNA16, ILND & 3D GAN & CSE-GAN with U-Net & TensorFlow, NVIDIA P100 &  & LUNA: 80.74\%, ILND: 76.36\% ; LUNA: 85.46\%, ILND: 82.56\% \\
  Tang et al. {\cite{tang2023artificial}} & 2023 & LUNA16 & NoduleNet & AI-based CT nodule diagnosis, Lung-RADS, PCA. & PyTorch, Nvidia GTX 1060 6GB &  & non-solid: 0.86; partially solid: 0.68; solid: 0.94; Class 1: 0.84, Class 3: 0.29, Class 5: 0.99 \\ 
  Luo et al. {\cite{luo2024pulmonary}} & 2024 & LUNA16 & RkcU-Net & Improved residual block & PyTorch, Python, NVIDIA P100 &  & DSC: 89.25\% ; Sensitivity: 88.48\% ; Precision: 90.04\% \\
  Bbosa et al. {\cite{bbosa2024mrunet}} & 2024 & LUNA16 & MRUNet-3D & MRUNet-3D: Multi-stride & PyTorch, NVIDIA 3060 &  & DSC: 83.47\% ; Sensitivity: 83.39\% ; Precision: 86.04\% \\
  Qiu et al. {\cite{qiu2023dual}} & 2023 & LIDC-IDRI, LUNA16 & Dual-task 3D U-Net & Dual-task region-boundary & PyTorch,NVIDIA RTX 3090 &  & LIDC-IDRI: 82.48 ± 8.17; LUNA16: 71.61 ± 14.17 \\
  Thangavel et al. {\cite{thangavel2024effective}} & 2024 & LIDC-IDRI, LUNA16 & T-Net, NASNet & T-Net, CenterNet, NASNet for seg & Keras, NVIDIA 930 mx &  & LIDC-IDRI: 99.07; LUNA16: 98.97 \\
  \midrule
  Chen et al. {\cite{chen2020lung}} & 2020 & LIDC-IDRI & LDDNet & LDDNet: Uses dense block, BN & PYDICOM, CV2 (OpenCV); 32GB &  & Accuracy: Over 99\% ; High segmentation accuracy and robustness \\
  Ni et al. {\cite{ni2022two}} & 2022 & LIDC-IDRI & Two-stage multitask U-Net & Coarse-to-fine 2-stage with 3D U-Net, MSU-Net. & PyTorch,3 NVIDIA GTX-1080 GPUs &  & Malignancy: 83.4\%; Margin: 81.4\%; Calcification: 92.4\% ; Accuracy 77.8\%; AUC 84.3\% \\
  Zhang et al. {\cite{zhang2022automatic}} & 2022 & LIDC-IDRI,TCIA, SHATMU & I-3D DenseUNet & Nested dense skip connection, TPS augmentation. & Keras/ TensorFlow, Linux, 6 NVIDIA 12GB &  & TCIA/LIDC: 0.8316, SHATMU: 0.8167 ; TCIA/LIDC: 0.9278, SHATMU: 0.9015 \\
  Usman et al. {\cite{usman2023mesaha}} & 2023 & LIDC-IDRI & MESAHA-Net & Multi-encoder structure & TensorFlow 2.0, NVIDIA & 5-fold cross-validation & 88.27\% ± 7.42\% ; 92.88\% ± 9.54\% ; 86.95\% ± 11.29\% \\
  Youssef et al. {\cite{youssef2023integrated}} & 2023 & LIDC-IDRI & 3D U-Net & MGRF model; 3D U-net for precise ROI segmentation. & PyTorch, Tesla V100 GPU (16GB) &  & Dice score: 93.64\% ± 5.20\% ; 93.30\% ± 07.72\% \\  
  Khanna et al. {\cite{khanna2020deep}} & 2020 & LUNA16, VESSEL12, HUG-ILD & Deep Residual U-Net & Residual U-Net, false-positive removal. & Keras/ Tensorflow, Intel i7, GTX 1060 (6GB) &  & LUNA16: 98.63\%; VESSEL12: 99.62\%; HUG-ILD: 98.68\% \\
  Nguyen et al. {\cite{nguyen2023manet}} & 2023 & LIDC-IDRI, LUNA16 & 3D UNet & UNet-based backbone, multi-branch attention. & NVIDIA Tesla P100 (16GB) &  & Consensus 3: 82.74 ± 8.11; Consensus 4: 83.61 ± 7.01 ; FROC sensitivity: 88.11\% \\
  \midrule
  Halder et al. {\cite{halder2020adaptive}} & 2020 & LIDC-IDRI & AMST & ASE: Adaptive filter, reduces false positives. & Siemens Somatom Spirit scanner &  & Min 0.9433; Max 0.9919; Avg 0.9872 ; 8-9 mm: 92.90\%, 9-10 mm: 93.55\%, 10-20 mm: 95.83\% \\
  Bhattacharyya et al. {\cite{bhattacharyya2023bi}} & 2023 & LUNA16 & DB-NET & DB-NET: Mish activation & PyTorch, NVIDIA GPU & 10-fold cross-validation & DSC: 88.89\% ; Sensitivity: 90.24\% ; Precision: 77.92\% \\
  Ma et al. {\cite{ma2024improved}} & 2024 & LUNA16, LNDb & Dig-CS-VNet & Dig-CS-VNet: Pixel threshold & PyTorch 1.7.0, RTX A5000x2 &  & LUNA16: 94.9\%, LNDb: 81.1\% ; LUNA16: 92.7\%, LNDb: 76.9\% \\
  \bottomrule
\end{tabular} 
\end{table*}

Key Innovations and Performance Comparisons: Recent AI-driven advancements in pulmonary nodule segmentation have significantly enhanced both accuracy and clinical utility. In 2020, Chen et al. \cite{chen2020lung} introduced the LDDNet model, achieving over 99\% segmentation accuracy on the LIDC-IDRI dataset, showing strong potential for reducing manual workload and improving consistency in clinical nodule identification. Khanna et al. \cite{khanna2020deep} similarly demonstrated the robustness of deep learning approaches with their deep residual U-Net, reaching a Dice Similarity Coefficient (DSC) of over 98\% on the LUNA16 dataset. In 2021, Dutande et al. \cite{dutande2021lncds} and Kadia et al. \cite{kadia2021r} presented a 2D-3D cascaded CNN and the R2U3D model, respectively, emphasizing the benefits of using multiple dimensionalities for capturing diverse features in nodule segmentation. In 2022, Zhang et al. \cite{zhang2022automatic} introduced the I-3D DenseUNet, excelling in segmenting complex tumor shapes, while Tyagi et al. \cite{tyagi2022cse} leveraged a CSE-GAN to overcome challenges posed by data scarcity, a common issue in medical imaging. Recent models have also focused on hybrid and multi-modal approaches: in 2023, Youssef et al. \cite{youssef2023integrated} combined deep learning with stochastic models for enhanced robustness, while Nguyen et al. \cite{nguyen2023manet} used a multi-branch attention mechanism to improve segmentation accuracy. In 2024, Bbosa et al. \cite{bbosa2024mrunet} and Sweetline et al. \cite{sweetline2024overcoming} developed MRUNet-3D and a multi-crop CNN, respectively, both demonstrating strong performance in segmenting small nodules—critical for the early detection necessary for effective lung cancer treatment.

Performance Evaluation via Dice Similarity Coefficient (DSC): Recent studies indicate significant progress in deep learning-based pulmonary nodule segmentation, particularly regarding the Dice Similarity Coefficient (DSC). Halder et al. \cite{halder2020adaptive} achieved a sensitivity of 94.88\% on the LIDC-IDRI dataset. Chen et al. \cite{chen2020lung} introduced LDDNet, which recorded over 99\% accuracy in lung parenchyma segmentation. Zhang et al. \cite{zhang2022automatic} reported an 83.16\% DSC with their I-3D DenseUNet on the LIDC dataset, while Liu et al. \cite{liu2024multiscale} reached a DSC of 87.50\% with their SCA-VNet on the LIDC-IDRI dataset. Sweetline et al. \cite{sweetline2024overcoming} developed a multi-crop CNN that achieved accuracies of 98.3\% and 98.5\% on the LUNA16 and LIDC-IDRI datasets, respectively. These advances, primarily based on U-Net variants, CNNs, and attention mechanisms, have significantly improved segmentation accuracy and reliability.

Model Adoption and Clinical Implications: The successful adoption of AI-driven segmentation models in clinical settings hinges on balancing accuracy with computational efficiency. While high-performing models like LDDNet and deep residual U-Net show great promise due to their high accuracy, they also present computational challenges that can hinder scalability, especially in resource-limited healthcare facilities. For instance, Zhang et al.'s \cite{zhang2022automatic} I-3D DenseUNet demonstrated excellent segmentation accuracy but may not be feasible for smaller healthcare centers due to its high computational demands. In contrast, Liu et al.'s \cite{liu2024multiscale} SCA-VNet offers a more balanced approach, achieving a DSC of 87.5\% while minimizing computational overhead, making it a more practical candidate for widespread clinical adoption. Additionally, reducing false positives remains a critical aspect of clinical deployment, as minimizing unnecessary procedures is vital for patient safety and healthcare efficiency. Approaches like attention mechanisms and ensemble learning, utilized in models such as MANet and SCA-VNet, enhance precision and sensitivity, ultimately making these models more reliable and suitable for real-world clinical use.

Future Directions and Clinical Potential: Future research in AI-driven pulmonary nodule segmentation should prioritize developing lightweight, computationally efficient models suitable for real-time use, especially in resource-limited settings. Such advancements could support integration into mobile devices or clinical workflows, improving access to diagnostic tools in underserved areas. Moreover, integrating segmentation models with other modalities, such as imaging, biomarkers, or genomic data, could enhance diagnostic accuracy and provide a comprehensive understanding of patient conditions. In conclusion, strides in AI-driven segmentation have improved accuracy and early lung cancer detection. Successful clinical integration will require balancing performance with practical deployment, focusing on efficiency, accessibility, and multimodal data to maximize impact on patient outcomes.

\subsection{Discussion on AI-Driven Advances in Lung Cancer Classification} 
Pulmonary nodule classification aims to distinguish between benign and malignant nodules and differentiate true nodules from imaging artifacts, which is crucial for guiding clinical decisions and reducing unnecessary interventions. Recent advancements in AI have significantly improved classification accuracy, enhancing patient care and resource utilization. To summarize these advancements, we reviewed 32 studies from 2020 to 2024, encompassing machine learning, deep learning, and hybrid approaches using datasets like LIDC-IDRI and LUNA16. Table ~\ref{table:6} highlights these studies, emphasizing their methodologies, strengths, limitations, and clinical relevance.

\begin{table*}
\scriptsize %
\setlength{\tabcolsep}{0.15cm} %
\renewcommand{\arraystretch}{1.5} %
\setlength{\heavyrulewidth}{1pt}
    \centering 
   \caption{Methods and their performance in pulmonary nodule classification on CT images.} 
\label{table:6}
\begin{tabular}{@{\extracolsep{\fill}} p{1.5cm} p{0.9cm} p{1.1cm} p{1.3cm} p{1.4cm} p{1.7cm} p{1.6cm} p{1.4cm} | p{3.1cm} @{}}
\toprule
  \makecell[l]{Authors} & \makecell[l]{Year} & \makecell[l]{Dataset} & \makecell[l]{Classification} & \makecell[l]{Technique/\\Network Type} & \makecell[l]{Main\\ Method} & \makecell[l]{Software/\\Hardware\\Utilized} & \makecell[l]{Validation/\\Test Method} & \makecell[l]{Model\\Performance} \\
  \midrule
\specialrule{0.1pt}{0pt}{0pt}
Savitha et al. {\cite{savitha2020holistic}} & 2020 & LIDC-IDRI & Nodule or non-nodule & DCNN +CRF & DCNN + CRF for extraction and classification & i5-7300-HQ, 16GB, GTX 1050 &  & ACC: Without CRF: 83\%; With CRF: 89.48\% \\
Naqi et al. {\cite{naqi2020lung}} & 2020 & LIDC-IDRI & Nodule or non-nodule & FODPSO; Geometric fit & Four phases: extraction, cla & MATLAB, Xeon 3.5 GHz &  & ACC: 96.90\% ; SEN: 95.60\% ; SPE: 97.00\% \\
Pinheiro et al. {\cite{de2020detection}} & 2020 & LIDC-IDRI & Nodule or non-nodule & CNNs & Swarm AI for CNN training & - &  & ACC: 93.71\% ; SEN: 92.96\% ; SPE: 98.52\% \\
Wang et al. {\cite{wang2021false}} & 2021 & LIDC-IDRI & Nodule or non-nodule & CS-LBP, ORT-EOH; H-SVMs & 3D feature extraction: CS-LBP, ORT-EOH & MATLAB, Xeon 3.5 GHz &  & Average1: 96.04\% ; Average2: 95.69\% ; Average3: 96.95\% \\
Naveen et al. {\cite{naveen2023approach}} & 2023 & LIDC-IDRI & Nodule or non-nodule & DS, RF, BPNN & DS, RF, BPNN for classification & - &  & solid: 98.00\%; part-solid: 93.68\%; non-solid: 97.20\% \\
Gugulothu et al. {\cite{gugulothu2023novel}} & 2023 & LIDC-IDRI & Nodule or non-nodule & SDMMT; U-Net;LTrP & Clas: CSDR-J-WHGAN & Python &  & CSDR-J-WHGAN: 97.11\%; GAN: 95.56\%; DCNN: 91.54\% \\
Chen et al. {\cite{chen2021ldnnet}} & 2021 & LUNA16; Kaggle DSB & Nodule or non-nodule & LDNNET: Dense-Block, BN, Dropout & LDNNET, Dense-Block: classification & 32GB RAM, 2.5 GHz CPU, GT 640M &  & LUNA16: 98.84\%; Kaggle DSB 2017: 99.95\% \\
Halder et al. {\cite{halder2023atrous}} & 2023 & LIDC-IDRI & benign or Malignant & Atrous CNN: ATCNN1P, ATCNN2PR & VGG-like structure: classification & GPU in Google Colab Pro &  & ATCNN2PR-1: 95.97\% ; ATCNN2PR-2: 95.84\% ; ATCNN2PR-3: 96.89\% \\
Sengodan et al. {\cite{sengodan2023early}} & 2023 & LIDC-IDRI & benign or Malignant & RCNN, Ensemble SVM & Ensemble SVM: classification & MATLAB, i7, 4 GB GPU & Train-Validation-Test Split & ACC: 98.53\% ; SEN: 99.30\% ; AUC: 0.98 ; SPE: 98.03\% \\
Guo et al. {\cite{guo20233d}} & 2023 & LUNA16 & benign or Malignant & 3D SAACNet + GBM & SAACNet + GBM: extraction and classification & PyTorch, 4 GTX 2080Ti &  & ACC: 95.18\%; SEN: 97.35\%; AUC: 97.70\%; SPE: 90.43\% \\
Sivakumar et al. {\cite{sivakumar2024efficient}} & 2024 & LUNA16, Kaggle DSB & benign or Malignant & ADBN, LightGBM & LightGBM: classification & TensorFlow, 16GB RAM &  & ACC: 99.87\% ; SEN: 99.75\% ; SPE: 99.42\% \\
Zhang et al. {\cite{zhang2021high}} & 2021 & Internal: 532pts & benign or Malignant & 3D CNN, SE-ResNet & 3D CNN: NSNs classification & TensorFlow, TITAN XP &  & AIS vs MIA-IAC: 81.9\% (CNN), 86.4\% (CNN + radio) \\
Harsono et al. {\cite{harsono2022lung}} & 2022 & LIDC-IDRI; Moscow & benign or Malignant & I3DR-Net & Transfer learning: I3D backbone, RetinaNet & PyTorch, Windows, Tesla P100 &  & Public-1: 94.12\%, Private-1: 65.90\%; Public-2: 81.84\%, Private-2: 70.36\% \\
Zhang et al. {\cite{zhang2022study}} & 2022 & LIDC-IDRI & Both & VGG16 & Seg: RW; Class: VGG16 + fused features & - &  & single VGG16: 0.930; multi VGG16: 0.975; multi-feature VGG16: 0.9681 \\
Cai et al. {\cite{cai2023impact}} & 2023 & LIDC-IDRI; HB; XZ & Both & 3D MaskRCNN, ResNet18-3D & Baseline models fine-tuned with local datasets & - &  & LIDC baseline: 0.846; HB: 0.813; XZ: 0.696 ; LIDC baseline: 0.837; HB: 0.849 \\
Gugulothu et al. {\cite{gugulothu2024early}} & 2024 & LIDC-IDRI & Both & CBSO; IFB; HDE-NN & Seg: CBSO; Classification: HDE-NN & Keras, 32 GB RAM, i5 &  & HDE-NN: 96.39\%; SVM: 91.68\%; ELM: 94.57\% ; HDE-NN: 95.25\%; SVM: 88.38\% \\
Raza et al. {\cite{raza2023lung}} & 2023 & IQ-OTH / NCCD & Both & EfficientNetB1 & Transfer learning: EfficientNet: class & Keras, TensorFlow; Tesla T4 &  & No augmentation: 98.64\%; With augmentation: 99.10\% \\
\midrule
Dodia et al. {\cite{dodia2022novel}} & 2022 & LUNA16 & Nodule or non-nodule & RFR V-Net; NCNet & RFR V-Net: seg; NCNet: clas & Keras; HPC setup &  & NCNet 3D: 98.21\%; NCNet 3D: 98.38\%; NCNet 3D: 98.33\% \\
Lei et al. {\cite{lei2020shape}} & 2020 & LIDC-IDRI & benign or Malignant & CNN, SAM, HESAM & SAM for shape \& margin analysis & PyTorch; 2 GTX 1080 Ti &  & HESAM: 99.13\%; SAM: 98.25\%; CAM: 96.51\% \\

  \bottomrule
\end{tabular}
\end{table*}
\begin{table*}
\scriptsize %
\setlength{\tabcolsep}{0.15cm} %
\renewcommand{\arraystretch}{1.7} %
\setlength{\heavyrulewidth}{1pt}
\begin{tabular}{@{\extracolsep{\fill}} p{1.5cm} p{0.9cm} p{1.1cm} p{1.3cm} p{1.4cm} p{1.7cm} p{1.6cm} p{1.4cm} | p{3.0cm} @{}}
\multicolumn{9}{l}{\small Table 6: (continued)}\\
\toprule

  \makecell[l]{Authors} & \makecell[l]{Year} & \makecell[l]{Dataset} & \makecell[l]{Classification} & \makecell[l]{Technique/\\Network Type} & \makecell[l]{Main\\ Method} & \makecell[l]{Software/\\Hardware\\Utilized} & \makecell[l]{Validation/\\Test Method} & \makecell[l]{Model\\Performance} \\
  \midrule
\specialrule{0.1pt}{0pt}{0pt}
Blanc et al. {\cite{blanc2020artificial}} & 2020 & SFR Data Challenge 2019 & Nodule or non-nodule & 3D U-NET; 3D Retina-UNET;SVM & SVM: classification & IBM Power AC922, Volta V100 GPU &  & (95\% CI:  84.83\% – 91.03\%); AUROC: 0.9058 \\
Fu et al. {\cite{fu2021fusion}} & 2021 & LIDC-CISB & benign or Malignant & 3D multi-classification; MLP & Mr-Mc and MLP for pathological types & Keras; Dual GTX1080ti (11GB) &  & Mr-Mc: 0.810; MLP: 0.887; Fusion: 0.906 ; Mr-Mc: 0.876; MLP: 0.980; Fusion: 0.950 \\
Amini et al. {\cite{amini2024fuzzy}} & 2024 & LIDC-IDRI; SPIE & benign or Malignant & FIG method & FIG for classification & - &  & Sub-band D2: 67.11\%; 8-orientation Gabor: 70.22\% \\
Wang et al. {\cite{wang2024towards}} & 2024 & LIDC-IDRI; Nanjing Univ. & benign or Malignant & Multi-task DL;3D nnU-Net; AAG and SAM & 3D nnU-Net: seg; ExPN-Net, AAG, SAM: classification & PyTorch, GPU & 5-fold cross-validation & LIDC: 95.5\%; In-house: 90.1\% ; LIDC: 100\%; In-house: 90.9\% ; LIDC: 0.992; In-house: 0.923 \\
Muzammil et al. {\cite{muzammil2021pulmonary}} & 2021 & LUNA16 & benign or Malignant & DCNN + SVM + AdaBoostM2 & Fusion: AlexNet, VGG-16, VGG-19 & MATLAB, i7-8550U, 8GB RAM &  & SVM + deep: 95.59\% ± 0.27\%; AdaBoostM2 + deep: 95.25\% \\
Wang et al. {\cite{wang2020fast}} & 2020 & LIDC-IDRI, LNUTCM & benign or Malignant & EOH, MSPLBP, NNCS & Seg:EOH, MSPLBP; DLSR (Class) & MATLAB, i7, 8GB RAM &  & ACC: 95.88\% ; AUC: 0.8149 (PR curves) \\
Wu et al. {\cite{wu2023self}} & 2023 & LIDC-IDRI; CQUCH & benign or Malignant & STLF-VA, 3D CNN & STLF-VA with self-supervised learning & Keras, Ubuntu, RTX TITAN &  & ACC: 85.30\% ; SEN: 86.80\% ; AUC: 0.9042 ; SPE: 83.90\% \\
Lin et al. {\cite{lin2020edicnet}} & 2020 & LIDC-IDRI; UCLA & Both & RetinaNet, ResNet 34, HSCNN & RetinaNet for detection; HSCNN for classification & PyTorch, Tesla V100 &  & Diameter, consistency, margin: 0.59, 0.74, 0.75 ; Mean AUC for malignancy: 0.89 \\
\midrule
Zhang et al. {\cite{zhang2022accurate}} & 2022 & LIDC-IDRI & Nodule or non-nodule & Radiomics; ML & Various models: classification & Python, CPU, GT 640M &  & RF + RFE: 0.9580 ; Best classifier (RF + RFE): 0.9893 \\
Suresh et al. {\cite{suresh2022nroi}} & 2022 & LIDC-IDRI & benign or Malignant & DCNN & Non-cancerous, malignant classification & MATLAB 2018b; GTX 960, 8 GB &  & ACC: 97.80\% ; SEN: 97.10\% ; AUC: 0.9956 ; SPE: 97.20\% \\
Gupta et al. {\cite{gupta2024texture}} & 2024 & LIDC-IDRI & benign or Malignant & SVM; LDA, KNN & Feature extraction: radiomics & MATLAB & 10-fold cross-validation & ACC: 91.3\% ; SEN: 90.0\% ; AUC: 0.96 ; SPE: 92.0\% \\
Zhai et al. {\cite{zhai2020multi}} & 2020 & LIDC-IDRI, LUNA16 & benign or Malignant & Multi-task CNN (MT-CNN) & MT-CNN for classification and reconstruction & PyTorch 1.3, Tesla V100, Python 3.7 &  & LUNA-16: 97.3\%; LIDC-IDRI: 95.59\% \\
Siddiqui et al. {\cite{siddiqui2023detection}} & 2023 & LIDC, LUNA16 & benign or Malignant & DBN, Gabor & Gabor filters with enhanced DBN & TensorFlow, 16 GB RAM &  & GF-DBN-SVM: 97.877\%; IGF-EDBN-SVM: 99.424\% \\

  \bottomrule
\end{tabular}
\end{table*}

Key Innovations and Performance Comparisons: Recent studies in pulmonary nodule classification have made significant strides in both model performance and clinical applicability. In the nodule versus non-nodule classification task, Savitha et al. \cite{savitha2020holistic} used a Deep Convolutional Neural Network (DCNN) combined with Conditional Random Fields (CRF) to reduce false positives, achieving a mean Intersection over Union (MIoU) of 0.911 and pixel accuracy of 89.48\% on the LIDC-IDRI dataset, with precision and recall of 0.95 each. Similarly, Naqi et al. \cite{naqi2020lung} utilized geometric fitting combined with deep learning, which reduced false positives to 2.8 with a sensitivity of 95.6\%, showcasing effective false positive control. Pinheiro et al. \cite{de2020detection} integrated swarm intelligence algorithms with CNNs to achieve 93.71\% accuracy on the LIDC-IDRI dataset, reducing training time by 25\%. Wang et al. \cite{wang2021false} improved multi-class nodule detection using 3D texture and edge features, reaching a sensitivity of 95.69\% and specificity of 96.95\%. Dodia et al.'s \cite{dodia2022novel} NCNet, combining V-Net and SqueezeNet, further enhanced sensitivity to 98.38\% while effectively managing false positives. These advancements highlight the diverse innovations targeting both high classification accuracy and the reduction of false positives, enhancing clinical decision-making capabilities.

In the benign versus malignant classification task, several methodologies have also emerged. Lei et al. \cite{lei2020shape} utilized SAM and HESAM methods for shape and edge analysis to reduce false positives. Amini et al. \cite{amini2024fuzzy} leveraged fuzzy information and texture features, achieving high accuracy, though with slightly lower sensitivity. Suresh et al. \cite{suresh2022nroi} achieved 97.8\% accuracy using a DCNN while managing false positives effectively, and Gupta et al. \cite{gupta2024texture} demonstrated that SVM models achieved 91.3\% accuracy on the LIDC-IDRI dataset. Zhai et al. \cite{zhai2020multi} used a Multi-task Convolutional Neural Network (MT-CNN) to reach an AUC of 97.3\% on the LUNA16 dataset, successfully reducing false positives. Similarly, Rahouma et al. \cite{rahouma2024automated} designed a genetic algorithm to optimize a 3D CNN, achieving 95.977\% accuracy, while Sivakumar et al. \cite{sivakumar2024efficient} demonstrated the potential of optimization algorithms with an accuracy of 99.87\%. These studies indicate progress in handling the complexities of benign versus malignant nodule classification, focusing on balancing sensitivity and false positive control.

For models that combine both nodule versus non-nodule and benign versus malignant classification, innovations in hybrid and transfer learning techniques have been instrumental. Lin et al. {\cite{lin2020edicnet}} developed EDICNet, integrating RetinaNet with hierarchical convolutional networks, achieving an AUC of 0.89 for malignant prediction. Cai et al. {\cite{cai2023impact}} enhanced classification performance through localized fine-tuning across multiple datasets, while Gugulothu et al. {\cite{gugulothu2024early}} combined chaotic bird optimization with an improved fish-swarm algorithm to boost accuracy and sensitivity. Raza et al. {\cite{raza2023lung}} demonstrated the effectiveness of Lung-EffNet, achieving 99.10\% accuracy on the IQ-OTH/NCCD dataset, addressing class imbalance with data augmentation techniques. These innovations highlight the critical role of advanced optimization and hybrid learning in achieving high accuracy across multiple classification tasks.

Regarding Computational Complexity: Recent studies indicate that improvements in pulmonary nodule classification often come with increased computational demands. Chen et al. \cite{chen2021ldnnet} reported higher complexity for their DenseNet model despite improved accuracy. Naqi et al. \cite{naqi2020lung} and Dodia et al. \cite{dodia2022novel} both enhanced model sensitivity but at the cost of increased computational burden. Lei et al. \cite{lei2020shape} similarly found their SAM and HESAM methods added significant complexity. In contrast, Suresh et al. \cite{suresh2022nroi} achieved high accuracy with relatively low computational overhead, emphasizing efficiency-focused design. These findings suggest that balancing performance with computational efficiency is key to making AI models viable for clinical use.

Model Adoption and Clinical Implications: The adoption of AI-driven classification models in clinical workflows requires balancing model performance with computational efficiency. High-performing models like DenseNet, employed by Chen et al. \cite{chen2021ldnnet}, improve performance but add computational complexity. Similarly, Naqi et al.'s \cite{naqi2020lung} FODPSO-based method successfully reduced false positives but at the cost of increased computational requirements. Conversely, Suresh et al. \cite{suresh2022nroi} employed a DCNN that maintained high accuracy with lower computational costs, making it more feasible for clinical integration. These examples illustrate the ongoing challenge of achieving high accuracy while maintaining computational feasibility, which is essential for broader clinical adoption, particularly in resource-limited settings.

Future Directions and Clinical Potential: Future research should focus on the development of computationally efficient models suitable for deployment in real-time clinical settings, particularly those with limited resources. Lightweight models capable of functioning on mobile devices could dramatically improve accessibility to high-quality diagnostic tools, especially in underserved areas. Moreover, the integration of imaging data with other modalities, such as genomic and biomarker information, presents a promising direction for enhancing diagnostic precision and providing a more comprehensive understanding of the patient’s condition. Ultimately, AI-driven classification models have made significant progress in distinguishing between benign and malignant nodules, enhancing early detection and reducing unnecessary interventions. To fully realize their clinical potential, continued focus on optimizing computational efficiency, accuracy, and data integration will be key to ensuring these tools benefit diverse healthcare environments.

Subtyping nodules enables more precise clinical decisions: we specifically discuss the subclassification of lung nodules—beyond the traditional benign/malignant dichotomy—to emphasize its clinical significance. Ground-glass nodules (GGNs), for example, are often early indicators of adenocarcinoma, whereas calcified nodules tend to be benign; recognizing these differences can directly influence follow-up intervals, biopsy decisions, and surgical planning. Recent studies demonstrate the feasibility of this finer-grained analysis: Naveen et al. (2023) report subtype-specific performance on LIDC-IDRI of 90.86\% for solid, 93.68\% for part-solid, and 97.06\% for non-solid nodules; Guo et al. (2023) and Sivakumar et al. (2024) achieve over 95\% accuracy on benign versus malignant classification, providing a strong foundation for multi-class extensions; and Zhang et al. (2021, 2022) incorporate segmentation and multi-task learning to fuse morphological and contextual features, further improving subtype discrimination. These advancements in multi-class classification and multi-task frameworks offer a more nuanced and clinically actionable approach to pulmonary CAD systems.

FDA clearance validates AI lung CAD for clinical use: we have added the following discussion of FDA-cleared AI systems to illustrate their clinical translation: Several AI systems for lung nodule detection and classification have obtained FDA clearance, demonstrating the readiness of deep learning–based CAD tools for clinical deployment. Optellum Lung Cancer Prediction (2021) leverages a deep learning–derived “lung cancer prediction score” trained on over 300,000 CT scans; a retrospective study in Radiology (2021) reported a 12–20 \% increase in pulmonologist diagnostic accuracy and a significant reduction in unnecessary follow-ups. Arterys Lung AI (2020) provides automated nodule detection, classification, and volumetric tracking via cloud-PACS integration, reporting robust performance on multi-institutional datasets for both solid and subsolid nodules. Although CT-based systems dominate, Lunit INSIGHT CXR (2021) earned FDA clearance for chest X-ray–based detection of ten thoracic abnormalities—including nodules—achieving an AUC of 0.974 in a clinical validation of over 100,000 images. Finally, VUNO Med-LungCT AI, cleared under Korea’s 510(k) pathway and pending FDA approval, has demonstrated > 90 \% sensitivity in multicenter Korean CT cohorts and is now integrated into early lung-cancer screening workflows. These examples underscore the importance of explainability, regulatory compliance, and seamless integration into clinical systems when translating deep learning methods from research to routine practice.

\begin{figure*}
  \begin{minipage}[t]{\linewidth}
    \centering
    \includegraphics[width=1\textwidth]{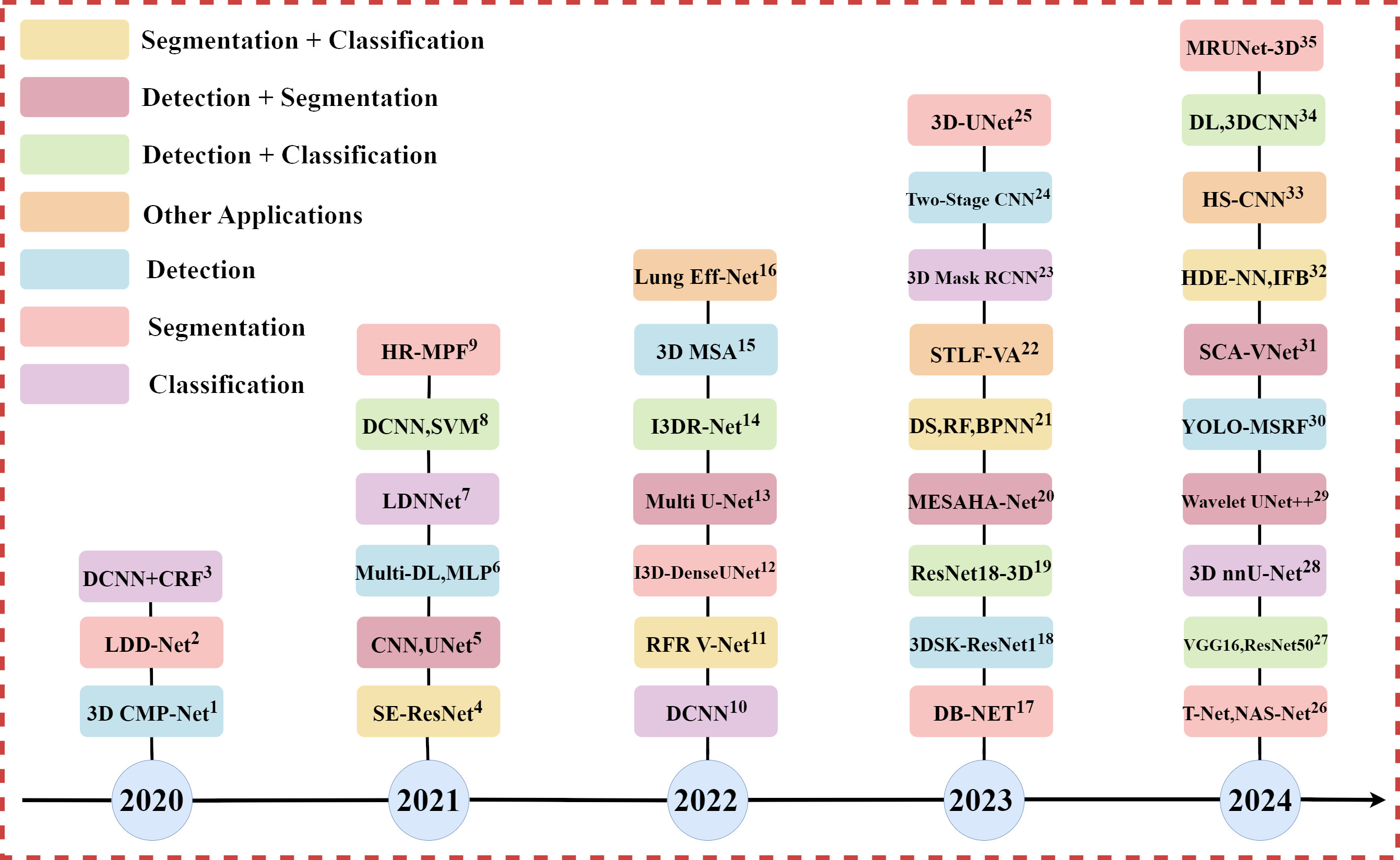}
    \caption{Timeline of AI-driven advancements in lung cancer detection, segmentation, and classification (2020–2024). Colors represent application areas: Segmentation + Classification (yellow), Detection + Segmentation (pink), Detection + Classification (green), Other Applications (orange), Detection (blue), Segmentation (red), and Classification (purple). Each node represents a key model or method with performance metrics, highlighting major breakthroughs. Superscript numbers next to models indicate their corresponding references: 1-3 \cite{song2020cpm, chen2020lung, savitha2020holistic}, 4-9 \cite{zhang2021high, osadebey2021three, fu2021fusion, chen2021ldnnet, muzammil2021pulmonary, zhu2021hr}, 10-16 \cite{suresh2022nroi, dodia2022novel, zhang2022automatic, ni2022two, harsono2022lung, zhang2022pulmonary, raza2023lung}, 17-25 \cite{bhattacharyya2023bi, zhang2023lungseek, cai2023impact, usman2023mesaha, naveen2023approach, wu2023self, cai2023impact, jain2023pulmonary, youssef2023integrated}, 26-35 \cite{thangavel2024effective, atiya2024classification, wang2024towards, agnes2024wavelet, wu2024yolo, liu2024multiscale, gugulothu2024early, lin2020edicnet, lin2024combined, bbosa2024mrunet}.}
    \label{fig:11}
  \end{minipage}
\end{figure*}

Through an in-depth analysis of the aforementioned studies on lung cancer detection, segmentation, and classification, {\hyperref[fig:11]{Fig. \ref*{fig:11}}} summarizes and presents the significant advancements made in these areas from 2020 to 2024. It comprehensively reflects the applications and performance improvements of deep learning models in pulmonary nodule analysis. The figure covers the evolution of single-task models (e.g., DCNN, 3D U-Net) and multi-task fusion models (e.g., ResNet18-3D, STLF-VA), clearly illustrating the trend of continuous optimization and integration of these technologies. With the introduction of advanced approaches such as YOLO-MSRF, I3DR-Net, and Wavelet U-Net++, AI-driven computer-aided diagnosis (CAD) systems have significantly enhanced detection sensitivity, segmentation precision, and classification accuracy. These systems demonstrate considerable potential, particularly in multi-task processing and complex network architectures. Notably, these models have achieved breakthroughs not only in improving early diagnosis and reducing misdiagnosis rates but also in providing robust support for real-world clinical applications. The developments outlined in {\hyperref[fig:11]{Fig. \ref*{fig:11}}} indicate that future trends will focus on multi-modal integration, automated feature extraction, and improving model generalization, thereby laying a solid technical foundation for early detection and accurate diagnosis of pulmonary nodules.

\section{Challenges and Future Prospects}
AI-driven advancements in lung cancer detection, segmentation, and classification have shown significant potential in improving diagnostic accuracy and efficiency. However, several challenges remain that hinder the full deployment and adoption of these technologies in clinical settings.

Data. A major challenge lies in the availability and quality of data. Although datasets like LIDC-IDRI and LUNA16 have been instrumental in advancing AI-based models for lung cancer analysis, these datasets are limited in scope and may not represent the full diversity of clinical cases. Furthermore, high-quality, annotated datasets are essential for training robust models, yet current datasets often suffer from annotation inconsistencies and a lack of standardization across institutions. The quality of CT images, including issues such as noise, motion artifacts, and variations in acquisition protocols, further complicates model training. These data challenges limit the generalizability of AI models and their robustness in real-world applications.

Algorithms. Selecting the appropriate algorithm for different tasks, such as detection, segmentation, and classification, remains a complex issue. Deep learning algorithms like CNNs, U-Net, and their variants have demonstrated strong performance, but they also present challenges in terms of model interpretability and computational cost. Many models function as "black boxes," making it difficult for clinicians to understand the decision-making process behind a model's predictions, potentially limiting clinical adoption. Moreover, there is a need for algorithms that are not only accurate but also computationally efficient to enable real-time diagnosis in resource-constrained environments.

Tasks. Current lung cancer CAD systems predominantly focus on well-defined tasks such as nodule detection, segmentation, and binary classification (benign vs. malignant). However, these tasks may not fully capture the complexity of lung cancer diagnosis. More nuanced tasks, such as predicting tumor growth or treatment response, require further exploration. Additionally, while most models focus on 2D image data from CT scans, there is an increasing need to integrate 3D data, multi-modal inputs (such as PET-CT), and even temporal data (such as follow-up scans) to better characterize the progression of lung cancer. This broader scope of tasks will require more sophisticated models capable of handling diverse and complex data types.

Tools. Although AI models have demonstrated impressive capabilities, there is still a significant gap in providing accessible tools for non-expert users, such as radiologists and clinicians, who may lack extensive programming knowledge. The development of user-friendly platforms that integrate classical machine learning and deep learning algorithms, offering ready-to-use tools for model training, evaluation, and interpretation, is crucial. Such platforms should allow clinicians to apply advanced AI models without needing to write complex code, thus reducing the time and technical barriers to AI adoption in clinical practice.

Generalizability and Validation. One of the major hurdles in implementing AI-driven CAD systems in real-world clinical settings is ensuring that models generalize well across different institutions and populations. Models trained on a specific dataset may not perform as expected when applied to data from a different hospital or region due to variations in imaging protocols, equipment, and patient demographics. Rigorous external validation across diverse datasets is required to address this issue. Additionally, cross-platform compatibility should be emphasized, ensuring that AI models can be effectively deployed across a range of hardware configurations, from high-performance servers to mobile devices.

In summary, AI-driven lung cancer diagnosis faces multiple challenges related to data, algorithms, tasks, tools, and generalizability, requiring reliable solutions to fully harness the potential of these technologies. Future research directions should focus on the following areas:

\begin{enumerate}
    \item \textbf{Building large, diverse datasets}: Expanding publicly available datasets with a broader range of imaging modalities (e.g., CT, PET, MRI) and clinical contexts (e.g., early-stage vs. late-stage lung cancer) to improve model robustness and generalizability.
    \item \textbf{Data augmentation and synthetic data}: Leveraging deep learning-based techniques such as GANs, to augment existing datasets and generate synthetic data, addressing issues of data scarcity and overfitting during model training.
    \item \textbf{Developing interpretable AI models}: Enhancing model transparency by integrating explainability methods, such as attention mechanisms or post-hoc interpretability tools, to help clinicians better understand AI predictions.
    \item \textbf{Creating efficient algorithms for real-time applications}: Designing lightweight, high-performance models that can be deployed in real-time diagnostic scenarios, particularly in resource-constrained environments such as rural clinics or mobile devices.
    \item \textbf{Expanding task scope}: Incorporating more complex tasks, such as longitudinal tumor tracking and multi-modal data integration, to provide a more comprehensive diagnostic tool for lung cancer management.
    \item \textbf{Building user-friendly AI platforms}: Developing zero-code or low-code platforms that allow clinicians to easily apply AI models, assess model performance, and generate predictions without requiring programming expertise.
    \item \textbf{Enhancing cross-institution generalizability}: Promoting external validation and cross-platform compatibility to ensure that AI models can generalize effectively across different clinical settings and patient populations.
\end{enumerate}

These future directions will be essential in overcoming the current challenges in AI-driven lung cancer diagnosis, making these technologies more accessible and reliable for clinical applications.

\section{Conclusion}
Deep learning has revolutionized medical image analysis, particularly in the early detection of lung cancer, where it surpasses traditional statistical methods in handling complex data. Despite notable advances in pulmonary nodule detection, segmentation, and classification using CNNs, RNNs, and GANs, challenges such as data scarcity and the interpretability of models remain. The reliance on large, annotated datasets limits the widespread clinical application of these models, while their "black box" nature raises concerns about transparency in medical decision-making. Future research should focus on enhancing model interpretability through explainable AI (XAI) techniques and overcoming data limitations with innovative solutions like GAN-based augmentation. Furthermore, integrating multimodal data—such as genomic and clinical information—into CAD systems and exploring hybrid models that combine deep learning with traditional machine learning could significantly improve diagnostic accuracy and clinical utility. By addressing these challenges, deep learning holds immense potential to further transform early lung cancer detection and patient outcomes. This review aims to provide valuable insights to guide future research in this evolving field.

\section*{CRediT authorship contribution statement}

\textbf{Guohui Cai:} Conceptualization, Investigation, Methodology, Writing – original draft.
\textbf{Ying Cai:} Investigation, Validation, Writing – original draft.
\textbf{Zeyu Zhang:} Writing – original draft.
\textbf{Yuanzhouhan Cao:} Methodology, Writing – review \& editing. 
\textbf{Lin Wu:} Investigation, Methodology.
\textbf{Daji Ergu:} Supervision.
\textbf{Zhibin Liao:} Investigation, Methodology.
\textbf{Yang Zhao:} Investigation, Supervision.

\section*{Declaration of competing interest}
The authors declare that they have no known competing financial interests or personal relationships that could have appeared to influence the work reported in this paper.

\section*{Data availability }
No data was used for the research described in the article. 

\section*{Acknowledgment }
This research has been supported by the National Natural Science Foundation of China (Grant No. 72174172) and the Scientific and Technological Innovation Team for Qinghai-Tibetan Plateau Research at Southwest Minzu University (Grant No. 2024CXTD20). We sincerely appreciate their valuable support, which made this work possible.

\end{document}